\documentclass[prd,10pt,aps,twocolumn,superscriptaddress,showpacs,nofootinbib,floatfix,amssymb,amsmath]{revtex4-1}
\usepackage{rotating}
\usepackage{multirow}

\usepackage{color}

\newcommand{\ext}{{\mathrm{ext}}}

\newcommand{\beqn}{\begin{eqnarray}}
\newcommand{\eeqn}{\end{eqnarray}}
\newcommand{\eq}[1]{(\ref{#1})}

\newcommand{\cL}{{\cal L}}

\newcommand{\cO}{{\cal O}}

\newcommand{\Z}{{\mathbb Z}}

\newcommand{\beqs}{\begin{subequations}}
\newcommand{\eeqs}{\end{subequations}}

\newcommand{\be}{\begin{equation}}
\newcommand{\ee}{\end{equation}}
\newcommand{\bal}{\begin{align}}
\newcommand{\eal}{\end{align}}
\newcommand{\bmt}{\begin{multline}}
\newcommand{\emt}{\end{multline}}

\begin{document}

\title{Electromagnetically superconducting phase of vacuum in strong magnetic field: structure of superconductor and superfluid vortex lattices in the ground state}

\author{M. N. Chernodub}\thanks{On leave from ITEP, Moscow, Russia.}
\affiliation{CNRS, Laboratoire de Math\'ematiques et Physique Th\'eorique, Universit\'e Fran\c{c}ois-Rabelais Tours,\\ F\'ed\'eration Denis Poisson, Parc de Grandmont, 37200 Tours, France}
\affiliation{Department of Physics and Astronomy, University of Gent, Krijgslaan 281, S9, B-9000 Gent, Belgium}
\author{Jos Van Doorsselaere}
\affiliation{Department of Physics and Astronomy, University of Gent, Krijgslaan 281, S9, B-9000 Gent, Belgium}
\author{Henri Verschelde}
\affiliation{Department of Physics and Astronomy, University of Gent, Krijgslaan 281, S9, B-9000 Gent, Belgium}

\begin{abstract}
Recently it was shown that vacuum in a background of strong enough magnetic field becomes an electromagnetic superconductor due to interplay between strong and electromagnetic forces. The superconducting ground state of the vacuum is associated with a spontaneous emergence of quark-antiquark condensates which carry quantum numbers of charged $\rho$ mesons. The $\rho$--meson condensate is an inhomogeneous structure made of the so-called $\rho$ vortices, which are parallel to the magnetic field axis. The condensation of the charged $\rho$ mesons induces a (much weaker) superfluid-like condensate with quantum numbers of the neutral $\rho^{(0)}$ mesons.  In this paper we show that the vortices in the superconducting condensate organize themselves in an equilateral triangular lattice similarly to an ordinary type--II superconductor. We show that each of these superconductor vortices is accompanied by three superfluid vortices and three superfluid antivortices made of the neutral $\rho$ meson condensate. The superconductor vortex overlaps with one of the superfluid vortices. The superposition of the superconducting and superfluid vortex lattices has a honeycomb pattern.
\end{abstract}

%Quantum chromodynamics, 12.38.-t
%Electromagnetic interactions, 13.40.-f
%Superconductivity, new topics in, 74.90.+n

\date{November 13, 2011}

\maketitle

\section{Introduction}

The behavior of quantum field theories in an extremely strong external magnetic field of hadronic scale has attracted growing interest of the scientific community. The chiral magnetic effect~\cite{Fukushima:2008xe}, which may take place in hot quark matter created in heavy-ion collisions~\cite{Kharzeev:2004ey}, provides a particularly interesting and experimentally observable example:  chirally--imbalanced matter generates electric current along the axis of the magnetic field~\cite{Vilenkin:1980fu}. The interest in the strong magnetic fields is far from being purely academic since extremely high magnetic fields may be generated in noncentral collisions of heavy ions~\cite{Fukushima:2008xe,Skokov:2009qp}, and they may have existed in the early moments of our Universe~\cite{Grasso:2000wj}.

Not only (dense) matter, but also a quantum vacuum may exhibit quite unusual properties in a sufficiently strong magnetic field background. A well-known related example is the magnetic catalysis~\cite{Klimenko:1991he,Gusynin:1994re,Gusynin:1995nb}, which implies, in particular, a steady enhancement of the chiral symmetry breaking in the vacuum of Quantum Chromodynamics (QCD) as the external magnetic field strengthens. Consequently, a strong background magnetic field affects drastically the phase structure of QCD vacuum~\cite{Gatto:2010pt,Mizher:2010zb,D'Elia:2010nq} and QCD matter~\cite{Preis:2011sp,Dexheimer:2011pz}

More recently it was shown that a sufficiently strong magnetic field of hadronic scale may cause the vacuum to behave as an inhomogeneous and anisotropic {\it electromagnetic} superconductor~\cite{Chernodub:2010qx,Chernodub:2011mc}. The superconductivity of, basically, empty space, is caused by a spontaneous creation of a (charged) $\rho$--meson condensate if the strength of the magnetic field exceeds the critical value
\beqn
B_c \simeq 10^{16} \, {\mathrm{Tesla}}
\qquad  {\mathrm{or}} \qquad
e B_c \simeq 0.6\,\mbox{GeV}^2\,.
\quad
\label{eq:Bc}
\eeqn 
The vacuum superconductivity is accompanied by a superfluid-like condensation of the neutral $\rho$ mesons~\cite{Chernodub:2010qx}.

The charged $\rho$ mesons -- or, better to say, quark-antiquark condensates with the quantum numbers of the $\rho$ mesons -- play a central role in the superconducting mechanism. The $\rho$ mesons are vector particles with anomalously high magnetic moment corresponding to the gyromagnetic ratio $g = 2$. One can argue that the anomalous magnetic moment provides a large negative contribution to the squared energy of the $\rho$ mesons in a background of strong magnetic fields. As the external magnetic field exceeds the critical value of the magnetic field~\eq{eq:Bc}, the energy becomes purely imaginary indicating a condensation of the $\rho$ mesons. The emergence of the electrically charged condensate implies, almost inevitably, electromagnetic superconductivity of the new vacuum ground state.~\cite{Chernodub:2010qx}

One can also argue that in the background of the strong magnetic field the charged vector mesons play a role of the Cooper pairs~\cite{Chernodub:2011tv}: the strong magnetic field makes the motion of the quarks essentially one dimensional because the electrically charged quarks may move only along the magnetic lines. In one spatial dimension a weak attraction between a quark (for example, ``up'' quark) and an antiquark (say, ``down'' antiquark) mediated by a (virtual) gluon inevitably leads to creation of a bound state, electrically charged vector meson (in our example, it is $\rho^+ \equiv u \bar d$ meson). The emergence of the bound states leads to lowering of the vacuum energy and, again, to condensation of the charged $\rho$ mesons.

The superconductivity of the vacuum in a strong magnetic field was first found in an effective bosonic model which describes the electrodynamics of the $\rho$ mesons~\cite{Chernodub:2010qx}. Later, the superconductivity effect was confirmed in the Nambu--Jona-Lasinio model~\cite{Chernodub:2011mc}. Signatures of this counterintuitive effect were also found in holographic approaches~\cite{Callebaut:2011ab,Ammon:2011je} and in numerical simulations of quenched lattice QCD~\cite{Braguta:2011hq}. 

Due to the anisotropic nature of the superconductivity (the vacuum superconducts only along the axis of the magnetic field) the Meissner effect is absent so that the $\rho$--meson condensate does not screen the external magnetic field~\cite{Chernodub:2010qx,Chernodub:2011tv}. Moreover, due to the anisotropic superconductivity the vacuum becomes a (hyperbolic) metamaterial which, electromagnetically, behaves similarly to diffractionless ``perfect lenses''~\cite{Smolyaninov:2011wc}.

This paper is devoted to a detailed study of the vortex contents of the superconducting ground state of the vacuum in a background of a strong magnetic field. In the superconducting phase the $\rho$--meson condensate forms an inhomogeneous periodic (lattice) structure made of the so-called $\rho$ vortices which are parallel to the magnetic field axis~\cite{Chernodub:2010qx}. The $\rho$ vortex -- which is a string-like topological defect in the $\rho$--meson condensate -- is a close analogue of the Abrikosov vortex in an ordinary superconductor.  In Section~\ref{sec:GL} we briefly outline a few basic properties of the Abrikosov vortex lattice in a mixed state of a type-II superconductor in the Ginzburg-Landau (GL) approach to superconductivity. Following this example, in Section~\ref{sec:rho} we discuss various properties of the $\rho$--vortex lattice such as the geometrical structure, inhomogeneous superconductivity and the structure of inhomogeneities -- including the vortex content -- of the neutral (superfluid) $\rho$--meson condensate. The last Section is devoted to our conclusions.

\section{Ginzburg-Landau model}
\label{sec:GL}

Consider the Abelian Higgs model which is a relativistic version of the GL model
\beqn
\cL(\phi,A) & = & - \frac{1}{4} F_{\mu\nu} F^{\mu\nu} + ({\mathfrak D}_\mu \phi)^* {\mathfrak D}^\mu \phi - V(\phi)\,, 
\label{eq:L:GL}\\
V(\phi) & = & - m^2 |\phi^2| + \frac{\lambda}{4} |\phi|^4\,,
\nonumber
\eeqn
where ${\mathfrak D}_\mu = \partial_\mu - i e A_\mu$ is the covariant derivative, $A_\mu$ is the electromagnetic field with the strength $F_{\mu\nu} = \partial_\mu A_\nu - \partial_\nu A_\mu$, and $\phi \equiv |\phi| \, e^{i \varphi}$ is the electrically charged scalar field with the mass parameter $m$ and the self-interaction coupling~$\lambda$.
The field $\phi$ -- which carries a unit ($e$) electric charge -- plays a role of a field of Cooper pairs (without loss of generality, we consider the singly-charged bosons $\phi$ instead of the usual doubly-charged bosons corresponding to the Cooper pairs of electrons).

In the condensed state, $m^2 > 0$, the ground state of the model~\eq{eq:L:GL} is characterized by a condensate $\langle \phi \rangle = \phi_0$ with $|\phi_0| = \sqrt{2 m^2/\lambda}$, 
while the photon $A_\mu$ and the scalar excitation $\delta \phi = \phi - \langle \phi \rangle$ acquire the following masses, respectively:
\beqn
m_{\phi} = \sqrt{2} m\,, 
\qquad 
m_{A} = \frac{2 e m}{\sqrt{\lambda}} \,.
\eeqn

We introduce an external magnetic field parallel to the $z$ axis, which is described by a ``symmetric'' gauge potential $A_\mu = A^\ext_\mu$,
\beqn
A_1 = - \frac{B_\ext}{2} x_2\,, \quad 
A_2 = \frac{B_\ext}{2} x_1\,, \quad 
A_3 = A_0 = 0\,, \quad 
\label{eq:Aext}
\eeqn
and ignore quantum fluctuations of the gauge fields and scalar fields, thus treating the problem at the classical level. We always assume that $B_\ext >0$ (and $e = |e|$) for the sake of simplicity.

We consider a type-II superconductor which corresponds to the following region of the parameters:
\beqn
\lambda > e^2\,.
\label{eq:typeII}
\eeqn
If the external magnetic field exceeds a certain (first) critical field, $B_\ext > B_{c_1}$, then the Abrikosov vortices are formed in the superconducting material. An elementary Abrikosov vortex carries the magnetic flux 
\beqn
\int d^2 x \, B(x) = \frac{2 \pi}{e}\,,
\eeqn
where the integration goes over the plane which is perpendicular to the vortex axis.

In the type-II regime the vortices repel each other and they form a regular periodic structure called ``the Abrikosov lattice''. If the strength of the magnetic field exceeds the second critical field, $B_\ext > B_{c_2}$ with
\beqn
B_c \equiv B_{c_2} = \frac{m^2}{e}\,,
\label{eq:Bc:GL}
\eeqn
then the superconductivity gets destroyed completely. 

We work in the regime quite close to the critical field~\eq{eq:Bc:GL}:
\beqn
B_\ext < B_c \,, \qquad |B_c - B_\ext| \ll B_c\,.
\label{eq:Bext}
\eeqn
The ground state is independent of the time $t$ and $z$ coordinates, so that the equations of motion of the system~\eq{eq:L:GL} can be written in the complexified form:
\beqn
%MCH D -> \mathfrak D in these equations:
\partial (B + e |\phi|^2) - 2 e \phi^\dagger {\mathfrak D} \phi & = & 0\,, 
\label{eq:GL:Eq1}\\
({\mathfrak D} {\bar {\mathfrak D}} + m^2 - e B) \phi -  \lambda |\phi|^2 \phi & = & 0\,,
\label{eq:GL:Eq2}
% end of corrections
\eeqn
with $z = x_1 + i x_2$,  ${\bar z} = x_1 - i x_2$, $\partial = \partial_1 + i \partial_2$,  ${\bar \partial} = \partial_1 - i \partial_2$, $A = A_1 + i A_2$,  ${\bar A} = A_1 - i A_2$,  $B \equiv F_{12} = - i({\bar \partial} A - \partial {\bar A})$/2, and the covariant derivatives are:
\beqn
{\mathfrak D} = \partial + \frac{e}{2} B_\ext z \,, 
\qquad
{\bar {\mathfrak D}} = {\bar \partial} + \frac{e}{2} B_\ext {\bar z}\,, 
\label{eq:mathfrak:D}
\eeqn

As the magnetic field increases towards $B_c$, the condensate $\phi$ diminishes and at the critical value of the field the condensate vanishes completely, $\phi (B_\ext = B_c) = 0$. Thus in the regime~\eq{eq:Bext}, the classical equations of motion \eq{eq:GL:Eq1} and \eq{eq:GL:Eq2} can be linearized. In the leading order we get the following equation for the condensate $\phi$:
\beqn
\bar {\mathfrak D} \phi & = & 0\,.
\label{eq:Dphi}
\eeqn
The solution $\phi = \phi(z)$ of this equation determines the magnetic field,
\beqn
B(z) & = & B_\ext + e \Bigl[|\phi(z)|^2 - \langle |\phi |^2 \rangle\Bigr]\,,
\label{eq:Bz:GL}
\eeqn
which is a certain function of the transversal  coordinates $x_1$ and $x_2$. The brackets $\langle \dots \rangle$ indicate a mean value in the transversal $(x_1,x_2)$ plane:
\beqn
\langle \cO \rangle = \frac{1}{{\mathrm{Area}}_\perp} \int d x_1 \int d x_2 \, \cO(x_1,x_2)\,,
\label{eq:plane:average}
\eeqn
where ${\mathrm{Area}}_\perp$ is the area of the transversal plane. An additive coordinate-independent term $\langle |\phi |^2 \rangle$ in the solution~\eq{eq:Bz:GL} is a normalization term which imposes the conservation of the magnetic flux coming through the transversal plane, 
\beqn
\langle B\rangle = B_\ext\,.
\label{eq:Bconservation}
\eeqn

Following Abrikosov, we choose a general solution of Eq.~\eq{eq:Dphi} in a form of a sum over lowest Landau levels~\cite{ref:Abrikosov}:
\beqn
\phi(z) = \sum_{n \in \Z} C_n h_n\Bigl(\nu, \frac{z}{L_B},\frac{{\bar z}}{L_B}\Bigr)\,,
\label{eq:phi:z:GL}
\eeqn
where 
\beqn
h_n(\nu, z, {\bar z}) = \exp\Bigl\{ - \frac{\pi}{2} \bigl(|z|^2 + {\bar z}^2\bigr) - \pi \nu^2 n^2 + 2 \pi \nu n {\bar z}\Bigr\}\,, \qquad
\label{eq:h:z}
\eeqn
were
\beqn
L_B = \sqrt{\frac{2 \pi}{ e B}}\,,
\label{eq:LB}
\eeqn
is the magnetic length and $\nu$ is an arbitrary parameter.

We anticipate that the solution corresponds to a periodic lattice structure, and thus we assume that the variable $n$ takes integer values in Eq.~\eq{eq:Dphi}, $n \in \Z$. The solution is parametrized by (arbitrary, in general) complex parameters $C_n$. In order to ensure a regular structure of the lattice, the coefficients $C_n$ are usually chosen in a periodic manner:
\beqn
C_{n+N} = C_n\,.
\label{eq:N:fold}
\eeqn
The solution with $N = 1$ defines the square lattice (with all $C_n$'s being equal). It corresponds to the original Abrikosov's solution~\cite{Abrikosov:1956sx}. 

In order to determine the optimal minimal-energy values of the vortex lattice parameters ($N$, $C_n$ and $\nu$) in the GL model one usually solves classical equations of motion. In this article we would like to use an explicit energy minimization over these parameters because this approach is more suitable for a {\it {nonlocal}} energy functional of the $\rho$-meson sector of the QCD vacuum.

The values of the coefficients $C_n$ in Eq.~\eq{eq:phi:z:GL} and the parameter $\nu$ in Eq.~\eq{eq:h:z} are determined by a minimization procedure of the action of the theory~\eq{eq:L:GL} in strong magnetic field. Since the action density is a function of the transversal coordinates $x_1$ and $x_2$, it it convenient to consider the energy density which is averaged over the transversal plane:

\beqn
\langle {\cal E} \rangle & = & \frac{1}{2} B_\ext^2 - (m^2 - e B_\ext) \langle |\phi|^2\rangle \nonumber \\
& & + \frac{e^2}{2} \langle |\phi|^2\rangle^2 + \frac{1}{2} \Bigl(\lambda - e^2 \Bigr) 
 \langle |\phi|^4\rangle\,.
\label{eq:GL:E}
\eeqn

Using the explicit form of the solution given by Eqs.~\eq{eq:phi:z:GL} and \eq{eq:h:z}, one can show that
\beqn
\langle |\phi|^2 \rangle & = & \frac{1}{\sqrt{2} |\nu|} \lim_{M \to \infty} \frac{1}{M} \sum_{n = - M/2}^{M/2} |C_n|^2\,, 
\label{eq:phi2:explicit}\\
\langle |\phi|^4 \rangle & = & \frac{1}{2 |\nu|} \lim_{M \to \infty} \frac{1}{M} \sum_{n_1 = - M/2}^{M/2} \sum_{n_2 \in \Z} \sum_{n_3 \in \Z}
e^{- \pi \nu^2 (n_2^2 + n_3^2)} \nonumber \\
& & C_{n_1+n_2} C_{n_1}^* C_{n_1+n_3} C_{n_1+n_2+n_3}^* \,.
\eeqn
In the case of the $N$-fold symmetry~\eq{eq:N:fold} one gets:
\beqn
\langle |\phi|^2 \rangle & = & \frac{1}{\sqrt{2} N |\nu|} \sum_{n = 0}^{N-1} |C_n|^2\,, 
\label{eq:phi2:N}\\
\langle |\phi|^4 \rangle & = & \frac{1}{2 N |\nu|} \sum_{n_1 = 0}^{N-1} \sum_{n_2 \in \Z} \sum_{n_3 \in \Z}
e^{- \pi \nu^2 (n_2^2 + n_3^2)} \nonumber \\
& & C_{n_1+n_2} C_{n_1}^* C_{n_1+n_3} C_{n_1+n_2+n_3}^* \,,
\label{eq:phi4:N}
\eeqn
where the $N$--fold symmetry~\eq{eq:N:fold} is assumed. 

In the Bogomolny limit of the couplings,
\beqn
\lambda = e^2\,,
\eeqn
the energy~\eq{eq:GL:E} depends only on the spatial average of the condensate squared $\langle |\phi|^2 \rangle$. 
At  $B_\ext < B_c$ the minimum of the energy density 
\beqn
{\cal E}_{\lambda= e^2}^{\mathrm{min}} = \frac{1}{2} (2 B_\ext - B_c) B_c\,,
\label{eq:E:min}
\eeqn
corresponds to the following value of the condensate:
\beqn
\langle |\phi|^2 \rangle_{\lambda= e^2}^{\mathrm{min}} = \frac{1}{e} (B_c - B_\ext)\,.
\label{eq:phi2:min}
\eeqn
Notice that in the Bogomolny limit an infinite amount of the lattice structures correspond to the same minimum of the energy functional~\eq{eq:E:min} because the energy minimum is determined only by the specific value of the condensate~\eq{eq:phi2:min}. The mean condensate defines  -- according to Eq.~\eq{eq:phi2:explicit} -- a mean value of the squared coefficients $C_n$, and does not determine their precise values. Therefore, in the Bogomolny limit the lattice is no more regular. The later fact nicely fits the observation that parallel Abrikosov vortices do not interact in the Bogomolny limit, so that a periodic lattice structure is unlikely to be formed. In the type--II regime~\eq{eq:typeII} the vortices repel each other so that the regular (periodic) vortex lattice is an anticipated structure of the ground state in this case.

Since we are working with a type--II superconductor~\eq{eq:typeII}, the last term in the energy density is always positive~\eq{eq:GL:E}. Therefore the global minimum of the energy~\eq{eq:GL:E} -- as a function of the parameter $N$ which determines the symmetry in Eq.~\eq{eq:N:fold} -- corresponds to a global minimum of a dimensionless quantity,
\beqn
\beta_{A} = \frac{\langle |\phi|^4 \rangle}{\langle |\phi|^2 \rangle^2}\,,
\label{eq:beta}
\eeqn
which is known as the Abrikosov ratio~\cite{ref:Abrikosov}. The Abrikosov ratio~\eq{eq:beta} 
in the ground state
is independent of the value of the external magnetic field.

For the simplest lattice of the square type, $N=1$, the Abrikosov ratio~\eq{eq:beta} is $\beta(N=1) = 1.180$ which is reached at $\nu = 1$. However, at $N=2$ the Abrikosov ratio (and, consequently, the energy) reaches its global minimum, $\beta(N=2) \approx 1.1596$, at the following parameters:
\beqn
N=2:\qquad C_1 = \pm i C_0\,, \qquad \nu = \frac{\sqrt[4]{3}}{\sqrt{2}} \approx 0.9306\,. \qquad 
\label{eq:N2:GL}
\eeqn
This minimal-energy periodic pattern corresponds to the equilateral triangular lattice (which is sometimes called ``hexagonal'' lattice). At higher odd values of $N$, the Abrikosov ratio is higher than the $N=2$ minimum (for example, $\beta_{N=3} = 1.167$) while at even values of $N$ the minimization converges to the two-fold pattern corresponding to the triangular lattice ($\beta_{N=2k} = \beta_{N=2}$ with $k \in \Z$). A nice review of the GL theory of type--II superconductors in magnetic field, and, in particular, a review of the vortex lattice structure, can be found in Ref.~\cite{ref:type-II:Review}.

\begin{figure}
\begin{center}
\begin{tabular}{ll}
\includegraphics[width=37mm, angle=0]{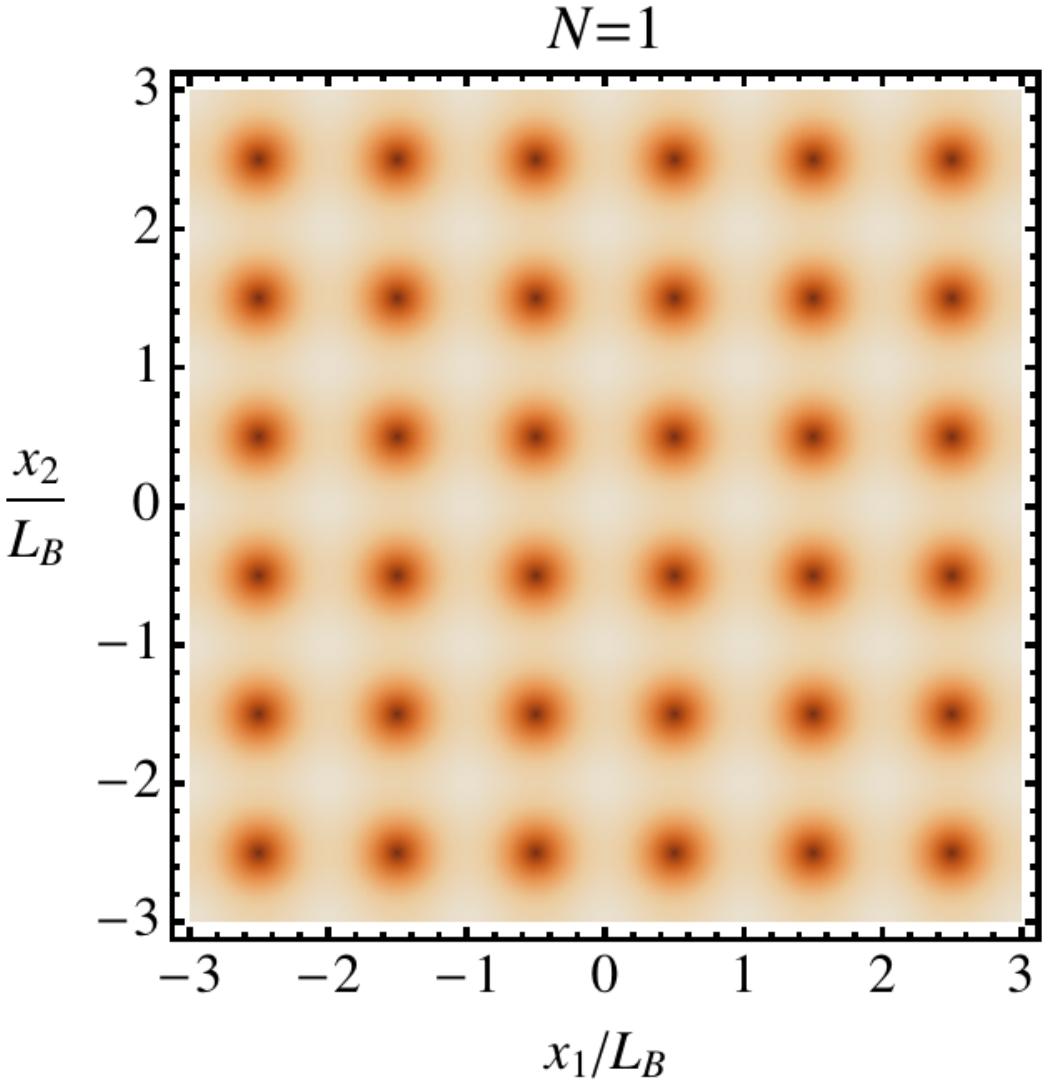} 
& \hskip 5mm
\includegraphics[width=37mm, angle=0]{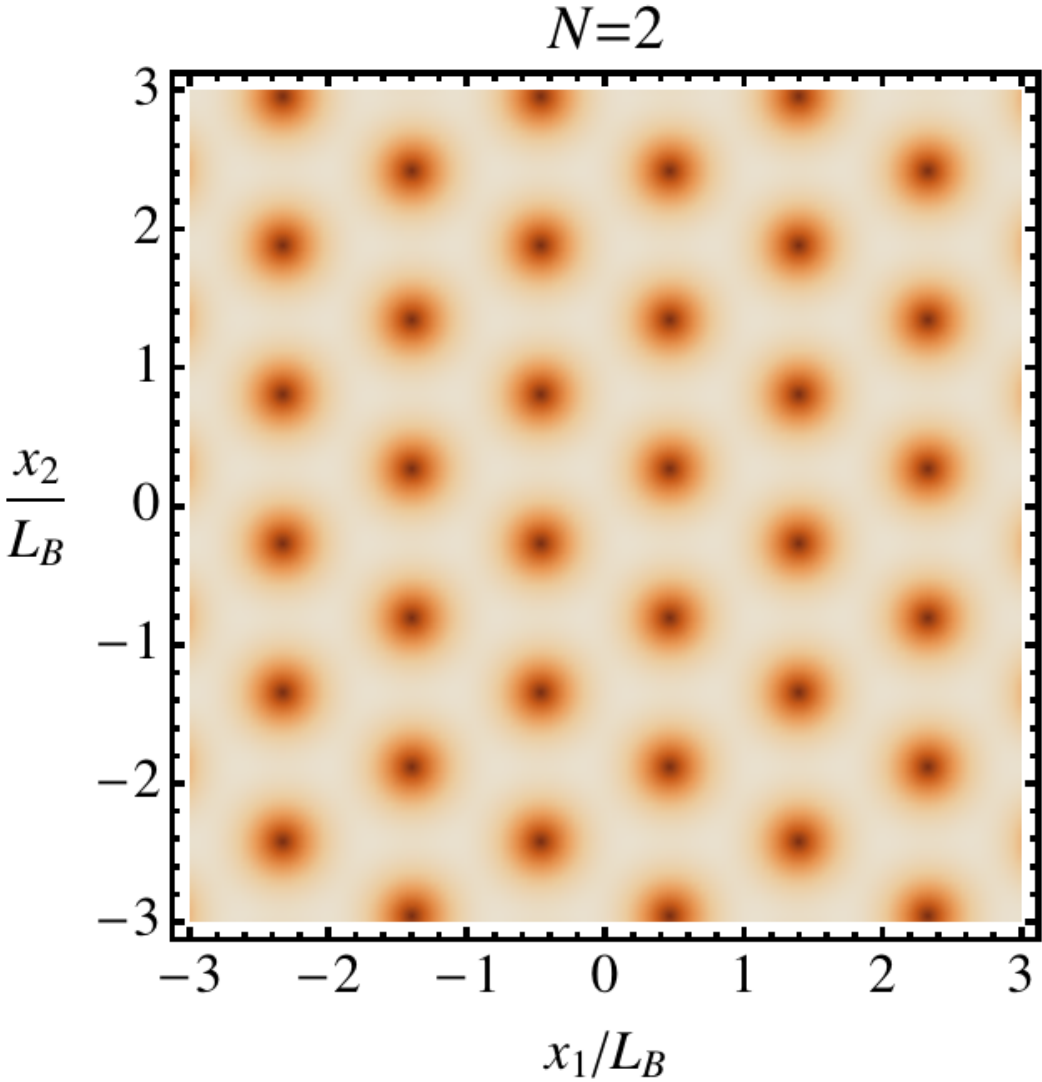}
\end{tabular}
\end{center}
\caption{Minimal-energy vortex lattices for $N=1$ (left) and $N=2$ (right) lattice symmetries in the GL model.}
\label{fig:GL:lattices}
\end{figure}

The square ($N=1$) and triangular ($N=2$) Abrikosov vortex lattices are visualized in Fig.~\ref{fig:GL:lattices}. The darker pointlike regions indicate the positions of the cores of the vortices characterized by a reduced value of the condensate $\phi = \phi(x_1,x_2)$. The condensate vanishes at the center of each vortex, $\phi = 0$, while the phase of the scalar field winds around a geometrical center of each vortex, $\arg \phi = \varphi$, where $\varphi$ is a two-dimensional azimuthal angle corresponding to a coordinate system centered at the vortex's origin.

According to Eq.~\eq{eq:Bz:GL}, the magnetic field is concentrated at the positions of the vortices, so that the distribution of the magnetic field is visually similar to Fig.~\ref{fig:GL:lattices}.

\section{Condensation of {\large $\mathbf{\rho}$} mesons in QCD}
\label{sec:rho}

The conventional superconductivity is driven by the condensate of the Cooper pairs. The superconductivity of vacuum in a strong magnetic field is guaranteed by quark-antiquark condensates which carry quantum numbers of (charged) $\rho$ mesons~\cite{Chernodub:2010qx}. Below we consider the electrodynamics of $\rho$ mesons in strong magnetic field.

\subsection{General properties}

\subsubsection{Electrodynamics of $\rho$ mesons}

The $\rho$ meson is a vector (spin-1) particle made of a light (up or down) quark and a light antiquark. A self-consistent quantum electrodynamics for the charged and neutral $\rho$ mesons is described by Djukanovic--Schindler--Gegelia--Scherer (DSGS) Lagrangian~\cite{Djukanovic:2005ag}:
\beqn
{\cal L} & = &-\frac{1}{4} \ F_{\mu\nu}F^{\mu\nu}
- \frac{1}{2} \ \rho^\dagger_{\mu\nu}\rho^{\mu\nu} + m_\rho^2 \ \rho_\mu^\dagger \rho^{\mu}
\label{eq:L:rho}\\
&& -\frac{1}{4} \ \rho^{(0)}_{\mu\nu} \rho^{(0) \mu\nu}+\frac{m_\rho^2}{2} \ \rho_\mu^{(0)}
\rho^{(0) \mu} +\frac{e}{2 g_s} \ F^{\mu\nu} \rho^{(0)}_{\mu\nu}\,,
\nonumber
\eeqn
which extends the vector meson dominance model~\cite{Sakurai:1960ju} with the Maxwellian $U(1)$ sector by adding all allowed interactions of both charged, $\rho_\mu \equiv \rho^- = (\rho^{(1)}_\mu - i \rho^{(2)}_\mu)/\sqrt{2}$ and $\rho^+_\mu = \rho^\dagger_\mu$, and neutral, $\rho^{(0)}_\mu$, mesons with the electromagnetic field $A_\mu$. The tensor quantities in \eq{eq:L:rho} correspond to various strength tensors,
\beqn
F_{\mu\nu} & = & \partial_\mu A_\nu-\partial_\nu A_\mu\,,
\label{eq:F}\\
{f}^{(0)}_{\mu\nu} & = & \partial_\mu \rho^{(0)}_\nu-\partial_\nu \rho^{(0)}_\mu\,,
\label{eq:f0}\\
\rho^{(0)}_{\mu\nu}& = & {f}^{(0)}_{\mu\nu}
- i g_s(\rho^\dagger _\mu \rho_\nu-\rho_\mu\rho^\dagger _\nu)\,,
\label{eq:rho0}\\
\rho_{\mu\nu} & = & D_\mu \rho_\nu - D_\nu \rho_\mu\,,
\label{eq:rho}
\eeqn
where the covariant derivative is
\beqn
D_\mu = \partial_\mu + i g_s \rho^{(0)}_\mu - i e A_\mu\,,
\eeqn
and $g_s \equiv g_{\rho\pi\pi} \approx 5.88$ is the $\rho \pi \pi$ vertex coupling.

The Lagrangian~\eq{eq:L:rho} respects the local $U(1)$ symmetry:
\beqn
U(1)_{\mathrm{e.m.}}: \quad
\left\{
\begin{array}{lcl}
\rho_\mu(x) & \to & e^{i \omega(x)} \rho_\mu(x)\,,\\
A_\mu(x) & \to & A_\mu(x) + \partial_\mu \omega(x)\,. \quad
\end{array}
\right.
\label{eq:gauge:invariance}
\eeqn

\subsubsection{The $\rho$-meson condensation due to strong magnetic field}

The last term of the DSGS Lagrangian~\eq{eq:L:rho} describes a nonminimal coupling of the $\rho$ mesons to the electromagnetic field which implies the anomalously large gyromagnetic ratio ($g = 2$) of the charged $\rho^\pm$ mesons. In  the background of a strong enough magnetic field a spin-one particle with  gyromagnetic ratio $g = 2$ should experience a tachyonic instability towards formation of a Bose-Einstein condensate. 

In a simple way the condensation can be explained as follows~\cite{Chernodub:2010qx}. A free charged relativistic spin-1 particle with the gyromagnetic ratio $g = 2$ and the mass $m$ has the following energy spectrum in a background of an external magnetic field $\vec B_\ext = (0,0,B_\ext)$:
\beqn
\varepsilon_{n,s_z}^2(p_z) = p_z^2+(2 n - 2 s_z + 1) eB_\ext + m^2\,.
\label{eq:energy:levels}
\eeqn
where $s_z = -1, 0, +1$ is the spin projection on the field's axis $\hat z \equiv {\hat x}_3$, $n\geqslant 0$ is a nonnegative integer number (which, together with $s_z$, labels the Landau levels), and $p_z$ is the particle momentum along the field's axis. The ground state (with $n=0$, $s_z = +1$, $p_z = 0$) has the following (squared) energy: 
\beqn
\varepsilon_0^2(B_\ext) = m^2 - e B_\ext\,.
\label{eq:E0}
\eeqn

The lowest energy of the charged $\rho$-meson in the external magnetic field becomes purely imaginary quantity if the magnetic field exceeds the critical value $B_c = m^2/e$. Exactly the same result, Eq.~\eq{eq:E0}, can be derived from the DSGS model~\eq{eq:L:rho} in a strong magnetic field background~\cite{Chernodub:2010qx}. The condensed state corresponds to the following $\rho$-meson wavefunction:
\beqn
\rho_1 = - i \rho_2 = \rho\,,
\qquad
\rho_0 = \rho_3 = 0\,.
\label{eq:wavefunction}
\eeqn

Thus, at $B_\ext > B_c$ the system experiences a  tachyonic instability towards the Bose-Einstein condensation of the spin-1 particles. A similar condensation effect -- a condensation of the $W$-bosons -- is suggested to happen in the standard model of electroweak interactions in a background with a much stronger magnetic field~\cite{Ambjorn:1988tm,Ambjorn:1988gb}. Another example is represented by a pure Yang-Mills theory, in which the gluons may condense in a {\it chromo}magnetic field background~\cite{Nielsen:1978rm}.

The condensation of the $\rho$ mesons in the vacuum takes place at the critical strength of the magnetic field~\eq{eq:Bc} defined by the $\rho$-meson mass $m_\rho = 775.5\,\mbox{MeV}$. The condensation of the charged excitation -- such as the $\rho$ meson -- should lead to an electromagnetic superconductivity of the vacuum.  We would like to stress that the superconducting state possesses the quantum numbers of the vacuum (i.e., the local electric charge density is zero at every point of the space--time). 

\subsubsection{Energy density of the ground state}

Let us consider the spontaneous condensation of the $\rho$ mesons in the background of the magnetic field~\eq{eq:Aext} in the vicinity of the phase transition: $B>B_c$ with $B - B_c {\ll} B_c$. 
Since the external field is taken to be slightly stronger than the critical value~\eq{eq:Bc}, the $\rho$-meson condensate is small, $|\rho| \ll m_\rho$, so that the equations of motion for $\rho$-meson electrodynamics~\eq{eq:L:rho} can be linearized. Following the example of the Ginzburg--Landau model described in Section~\ref{sec:GL}, we consider static $x_3$--independent solutions which may, however, be inhomogeneous in the transversal $(x_1,x_2)$ plane. 

The classical equations of motion in the magnetic field background are quite complicated and we refer an interested reader to Ref.~\cite{Chernodub:2010qx} for a detailed discussion. It turns out that the equation for the $\rho$-meson condensate~\eq{eq:wavefunction} in the overcritical magnetic field ($B \gtrsim B_c$) is similar to the one for the Cooper pair condensate in the subcritical magnetic field ($B \lesssim B_c$) in the Ginzburg--Landau model:
\beqn
{\mathfrak D} \rho \equiv (\partial -\frac{e}{2} B_\ext z) \rho = 0\,,
\label{eq:cDrho0}
\eeqn
where the covariant derivative is given in Eq.~\eq{eq:mathfrak:D} and the complex wavefunction
$\rho = \rho(z)$ is defined in Eq.~\eq{eq:wavefunction}

Solutions of Eq.~\eq{eq:cDrho0} are inhomogeneous in the transversal $(x_1,x_2)$ plane, and these inhomogeneities induce condensation of the neutral mesons, $\rho^{(0)} = \rho^{(0)}_1 + i \rho^{(0)}_2$:
\beqn
\rho^{(0)}(x_\perp) = \frac{2 i g_s}{- \partial^2_\perp + m^2_0} \partial |\rho|^2\,,
\label{eq:rho0:rho}
\eeqn
with $\rho^{(0)}_0 = \rho^{(0)}_3 = 0$. In Eq.~\eq{eq:rho0:rho} $\partial^2_\perp \equiv \partial^2_1 + \partial^2_2$,
\beqn
\frac{1}{- \partial^2_\perp + m^2_0}(x_\perp) = \frac{1}{2 \pi} K_0(m|x_\perp|)\,,
\eeqn
is the two-dimensional Euclidean propagator of a scalar particle with the mass of the neutral $\rho^{(0)}$ meson,
\beqn
m_0 \equiv m_{\rho^{(0)}} = m_\rho \Bigl(1 - \frac{e^2}{g_s^2}\Bigr)^{-\frac{1}{2}}\,,
\label{eq:m:rho0}
\eeqn
and $K_0$ is a modified Bessel function.

The magnetic field is also an inhomogeneous function of the transversal coordinates,
\beqn
B(x_\perp) = B_\ext + \frac{2 e m_0^2}{- \partial_\perp^2 + m^2_0} \Bigl[|\rho(z)|^2 - \langle{|\rho|^2}\rangle\Bigr] \,,
\qquad
\label{eq:B:solution}
\eeqn
where the last term, defined in Eq.~\eq{eq:plane:average}, guarantees the conservation of the magnetic flux~\eq{eq:Bconservation}. Equation \eq{eq:B:solution} is a $\rho$-meson analogue of the GL relation~\eq{eq:Bz:GL}.

Notice that in the vacuum subjected to the strong magnetic field, the neutral condensate~\eq{eq:rho0:rho} and the magnetic field~\eq{eq:B:solution} depend on the (charged) $\rho$--meson
condensate {\it nonlocally} contrary to the {\it local} relation between the magnetic field and the Cooper pair condensate~\eq{eq:Bz:GL} in the GL model for the ordinary superconductivity.

The energy density of the vacuum in the presence of the $\rho$-meson condensate is given by the following formula:
\beqn
\langle {\cal E} \rangle \equiv \langle T_{00} \rangle & = & \frac{1}{2} B_\ext^2 + 2 (m^2_\rho - e B_\ext) \langle |\rho|^2\rangle 
+ 2 e^2 \langle |\rho|^2\rangle^2  \nonumber \\
& & + 2 \bigl(g_s^2 - e^2\bigr) \left\langle \vert \rho\vert^2\frac{m^2_0}{ - \Delta + m^2_0}\vert \rho\vert^2\right\rangle \,,
\label{eq:E:rho}
\eeqn
were $T_{\mu\nu}$ is the energy-momentum tensor,
\beqn
T_{\mu\nu} = 2 \frac{\partial {\cal L}}{\partial g^{\mu\nu}} - {\cal L}\, g_{\mu\nu}\,.
\eeqn
corresponding to the DSGS model~\eq{eq:L:rho}, and the brackets $\langle \dots \rangle$ indicate the average over the transversal $(x_1,x_2)$ plane~\eq{eq:plane:average}. Notice that contrary to the GL model~\eq{eq:GL:E}, the energy density of the $\rho$-meson condensate~\eq{eq:E:rho} contains a nonlocal positive definite ($g_s \gg e$) quartic term.

\subsection{Physical properties of the ground state}

\subsubsection{Superconductor vortices: the lattice structure}

Following our experience in the GL model we represent the solution of the $\rho$-meson condensate in a manner similar to Eq.~\eq{eq:phi:z:GL}, 
\beqn
\rho(z) = \sum_{n \in \Z} C_n h_n\Bigl(\nu, \frac{z}{L_B},\frac{{\bar z}}{L_B}\Bigr)\,,
\label{eq:rho:z}
\eeqn
where the function $h_n$ is given in Eq.~\eq{eq:h:z} and $L_B$ is the magnetic length~\eq{eq:LB}. The coefficients are assumed to obey the $N$--fold symmetry~\eq{eq:N:fold}. 

The quadratic term $\langle |\rho|^2 \rangle$ in the energy density~\eq{eq:E:rho} can be evaluated with the help of Eq.~\eq{eq:phi2:N}, while the local quartic term $\langle |\rho|^4 \rangle$ is absent in Eq.~\eq{eq:E:rho}. Instead, the condensate $\rho$ is stabilized by a nonlocal quartic term, which is proportional to the following nonlocal functional:
\beqn
Q[\rho]  = \left\langle \vert \rho \vert^2\frac{m^2_0}{ - \Delta + m^2_0}\vert \rho\vert^2\right\rangle\,,
\label{eq:Q:rho}
\eeqn
which can conveniently be represented as follows:
\beqn
Q[\rho] = \frac{1}{\mathrm{Area}_\perp} \int \frac{d^2 k}{(2 \pi)^2} \, \frac{m^2}{k^2 + m^2_0} |q(k;\rho)|^2\,,
\label{eq:Q}
\eeqn
where
\beqn
q(k;\rho) = \int d x_1 \int d x_2\,  e^{i k_1 x_1 + i k_2 x_2} |\rho(x_1,x_2)|^2\,,
\label{eq:q:fun}
\eeqn
with $q^*(k;\rho) \equiv q(-k;\rho)$ and $\rho(x_1,x_2) \equiv \rho(x_1 + i x_2)$.

Let us shift the momentum, $k \to k/L_B$, and the coordinate, $z \to L_B z$, so that the new variables $k$ and $z$ are now dimensionless quantities. Then quantity~\eq{eq:q:fun} takes the following form
\beqn
q(k;\rho) = L_B^2 \sum_{n_1 \in \Z} \sum_{n_2 \in \Z} C_{n_1} C^*_{n_2} H_{n_1,n_2}(k_1,k_2)\,,
\label{eq:q:rho}
\eeqn
where
\beqn
H_{n_1,n_2}(k_1,k_2) & = & \int d^2 z \, e^{i k \cdot z} \, h_{n_1}(z,\bar z) h_{n_2}({\bar z},z) \nonumber \\
& = & \sqrt{2} \pi \delta\Bigl(k_2 - 2\pi \nu (n_1 - n_2)\Bigr) e^{- \frac{\pi}{2} \nu^2 (n_1 - n_2)^2}  \nonumber \\
& & \cdot \exp\left\{i k_1 \nu \frac{n_1 + n_2}{2} - \frac{k_1^2}{8 \pi} \right\} \!,
\eeqn
with  $k \cdot z = k_1 x_1 + k_2 x_2$.

\begin{table*}[htb]
\begin{tabular}{c|c|c|c|c|c|c|c|c|l}
$N$ & $\beta_\rho$ & $\sqrt[4]{-\delta {\cal E}}$, MeV & $\nu$ & $C_0$, MeV & $C_1$ & $C_2$ &  $C_3$ &  $C_4$ & type\\
\hline
   1  &     1.02393   &    26.7334  & 1            &  10.9516 & & & & & square  \\
   2  &     1.01920   &    26.7656  & $\sqrt[4]{3}/\sqrt{2}$ &  10.5977 & $i C_0$ & & & & hexagonal   \\
   3  &     1.02056   &    26.7567  & 0.95581 &  10.9744 & $C_0\, e^{i\varphi_3}$ &  $C_0\, e^{i\varphi_3}$ & & & parallelogrammic   \\
%MCH correct value "1.01920" is inserted at N=4 row
   4  &     1.01920   &    26.7656  &$\sqrt[4]{3}/\sqrt{2}$ &  10.5977 & $C_0$  & $-C_0$  & $-C_0$ & & hexagonal   \\
   5  &     1.01954   &    26.7633  & 0.93883 &  10.9773 & $C_0 \, e^{-4i\varphi_5}$  & $C_0 \, e^{-2i\varphi_5}$  & $C_0 \, e^{-4i\varphi_5}$  & $\quad C_0 \quad$  & parallelogrammic   \\
\end{tabular}
\caption{The parameters of the $\rho$--meson lattices at $B_\ext = 1.01 B_c$: the dimensionless ratio $\beta_\rho$, Eq.~\eq{eq:beta:rho}, the condensation energy per unit volume, $\delta {\cal E}$, Eq.~\eq{eq:E:cond}, and the structure constants $C_n$ for the first five lattice types $N=1,\dots,5$ (here $\varphi_k = 2 \pi/k$). Examples of certain lattices are shown in Fig.~\ref{fig:rho:lattices}. The global minimum in energy is reached at $N=2$.}
\label{tbl:rho}
\end{table*}
\begin{figure*}[htb]
\begin{center}
\begin{tabular}{cccc}
\includegraphics[width=39mm, angle=0]{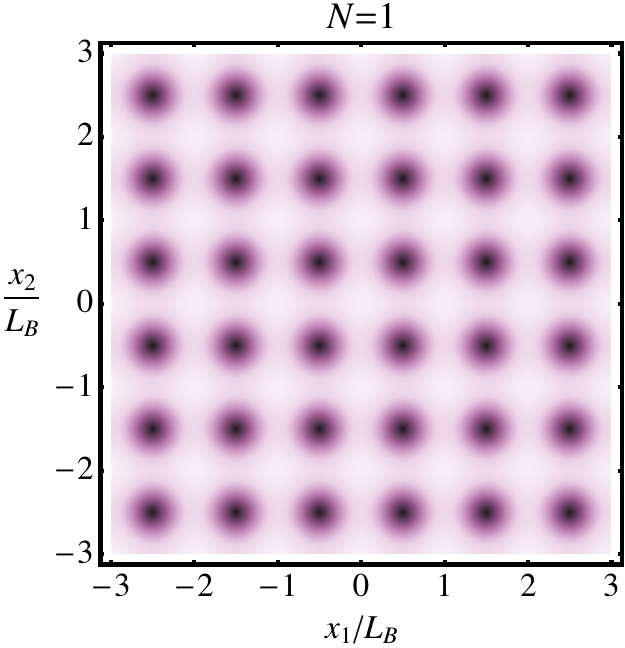}
& \hskip 5mm
\includegraphics[width=39mm, angle=0]{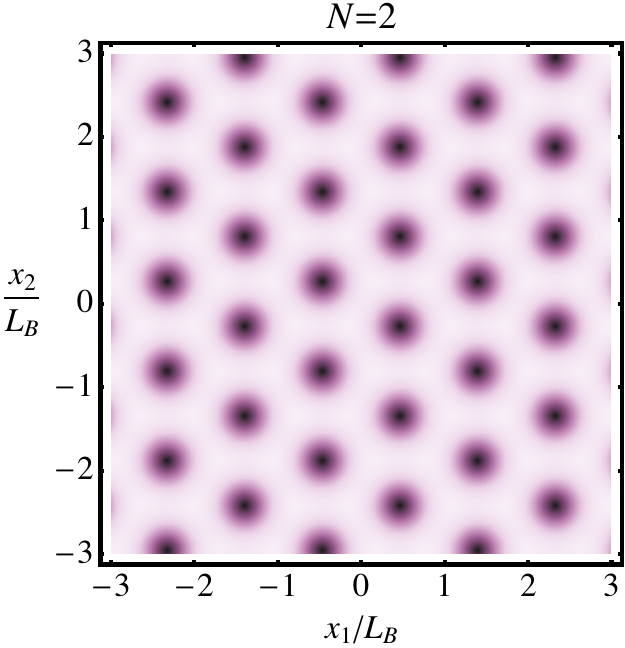} 
& \hskip 5mm
\includegraphics[width=39mm, angle=0]{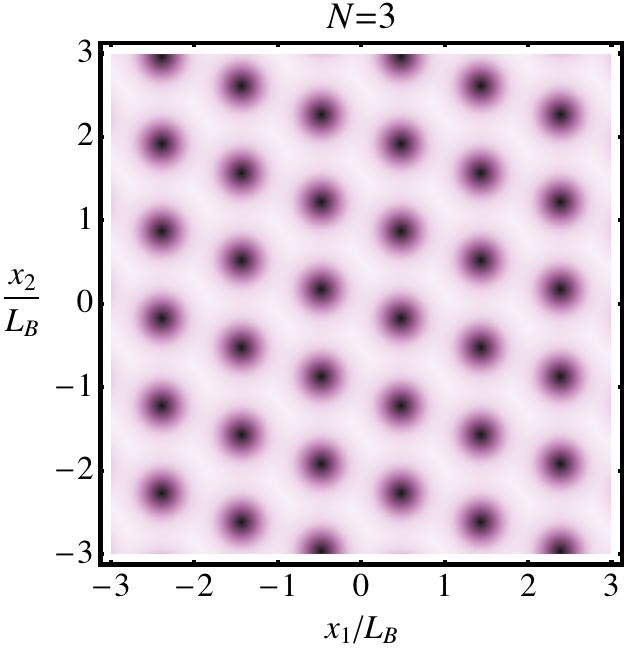}  
& \hskip 5mm
\includegraphics[width=39mm, angle=0]{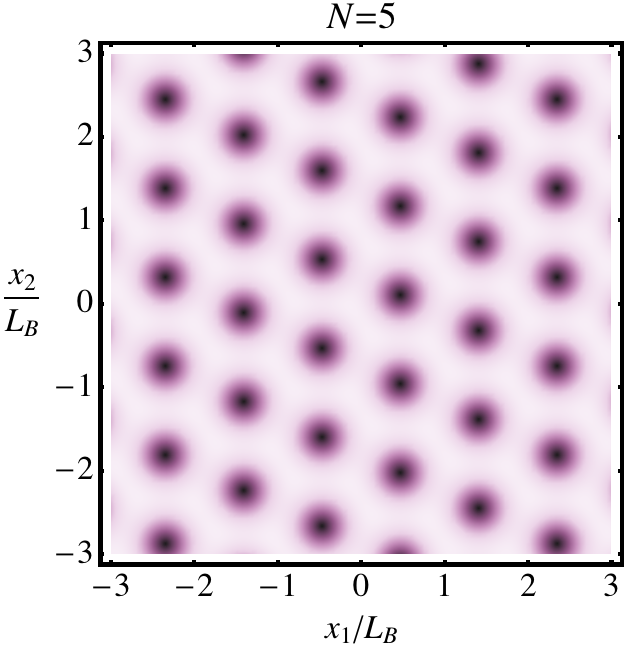}\\[0mm]
\hskip 5mm (a) & \hskip 11mm (b) & \hskip 11mm (c) & \hskip 11mm (d)
\end{tabular}
\end{center}
\caption{The density plots of the superconducting density in the $\rho$--meson vortex lattices with minimal energy at fixed lattice type (a) $N = 1$,  (b) $N = 2$,  (c) $N = 3$,  and (d) $N = 5$ at $B =1.01 B_c$. The corresponding magnetic length~\eq{eq:LB} is $L_B = 0.634\,\mathrm{fm}$. The darker regions corresponds to positions of the $\rho$ vortices where the superconducting density is suppressed.}
\label{fig:rho:lattices}
\end{figure*}

Next, we substitute Eq.~\eq{eq:q:rho} into Eq.~\eq{eq:Q}, 
\beqn
& & Q[\rho] =  \frac{1}{\mathrm{Area}_\perp}  \prod_{s=1}^4 \sum_{n_s \in \Z} C_{n_1} C_{n_2}^* C_{n_3} C_{n_4}^* \nonumber\\
& & \int \frac{d^2 k}{(2 \pi)^2} \, \frac{1}{k^2 + m^2}  H_{n_1,n_2}\bigl(k\bigr) H_{n_3,n_4}\bigl(-k\bigr) \,,
\label{eq:phi4:1:nonloc}
\eeqn
and take into account the following relation: 
\beqn
& & 
%MCH removed: \int d k_1 \, 
\delta\Bigl(k_2 - 2\pi \nu (n_1 - n_2)\Bigr) \, \delta\Bigl(k_2 + 2\pi \nu (n_3 - n_4)\Bigr) \nonumber \\
& = & \frac{N_2}{2 \pi |\nu|} \delta_{n_1 - n_2 + n_3 - n_4, 0}\, \delta\Bigl(k_2 - 2\pi \nu (n_1 - n_2)\Bigr)\,,
\eeqn
where $N_2$ is the number of elementary cells in $x_2$ direction of the Abrikosov lattice [notice that $\delta(x)$ is the Dirac $\delta$ function while $\delta_{n,0}$ is the Kronecker symbol]. Next, we make the following shifts of the integer-valued variables: $n_1 \to n_1 + n_2$, $n_3 \to n_3 + n_4$, and $n_4 \to n_1 - n_2 + n_3$. The result is
\beqn
Q[\rho] & = & \frac{1}{2} \frac{N_2\, L_B^2}{\mathrm{Area}_\perp} \sum_{n_1 \in \Z} \sum_{n_2 \in \Z} \sum_{n_3 \in \Z}
f_{n_2,n_3} (m_0 L_B, \nu) \nonumber \\
& & C_{n_1+n_2} C_{n_1}^* C_{n_1+n_3} C_{n_1+n_2+n_3}^* \,,
\label{eq:phi4:2:nonloc}
\eeqn
where the matrix $f_{n_2,n_3}$ is given by the following relation:
\beqn
f_{n_2,n_3} (\mu,\nu) = \frac{\mu^2}{|\nu|} \, \int\limits_{- \infty}^\infty \frac{d k}{2 \pi} \, \frac{e^{ - \pi \nu^2 n_2^2 - k^2/(4 \pi) + i k \nu n_3} }{k^2 + 4 \pi^2 \nu^2 n_2^2 + \mu^2}\!. 
\quad
\label{eq:f:fun}
\eeqn

We expect that in the infinite-mass limit, $m_0 \to \infty$, the functional~\eq{eq:Q:rho} should converge to the local quantity,
\beqn
\lim_{m_0 \to \infty} Q[\rho]  = \left\langle \vert \rho \vert^4\right\rangle\,.
\eeqn
And, indeed, in  the limit $\mu \equiv m L_B \to \infty$ one gets for the matrix~\eq{eq:f:fun} the following expression:
\beqn
\lim_{\mu \to \infty} f_{n_2,n_3}(\mu) = e^{- \pi \nu^2 (n_2^2 + n_3^2)}/|\nu|\,,
\label{eq:f:limit}
\eeqn 
so that the nonlocal functional~\eq{eq:phi4:nonloc} converges to the local expression~\eq{eq:phi4:N}, as expected. The function $f_{n_2,n_3}$ is not symmetric under discrete $\pi/2$ rotations in the $(n_2,n_3)$ plane at small values of the dimensionless parameter $\mu$. However, as the mass parameter $\mu$ increases, the function $f$ approaches the symmetric Gaussian function~\eq{eq:f:limit}.

Finally, taking into account the $N$--fold periodicity~\eq{eq:N:fold} of the coefficients $C$, one can rewrite Eq.~\eq{eq:phi4:2:nonloc} as follows:
\beqn
Q[\rho] & = & \frac{1}{2 N} \sum_{n_1 = 0}^{N-1} \sum_{n_2 \in \Z} \sum_{n_3 \in \Z}
f_{n_2,n_3} (m_0 L_B,\nu) \nonumber \\
& & C_{n_1+n_2} C_{n_1}^* C_{n_1+n_3} C_{n_1+n_2+n_3}^* \,.
\label{eq:phi4:nonloc}
\eeqn

The structure of the energy functional~\eq{eq:E:rho} indicates that an analogue of the Abrikosov ratio~\eq{eq:beta} in the case of the $\rho$--meson condensation is as follows:
\beqn
\beta_{\rho} = \left\langle \frac{\vert \rho \vert^2}{\langle |\rho|^2 \rangle} \frac{m^2_0}{ - \Delta + m^2_0}\frac{\vert \rho \vert^2}{\langle |\rho|^2 \rangle} \right\rangle
\equiv \frac{Q[\rho]}{\langle |\rho|^2 \rangle^2} \,.
\label{eq:beta:rho}
\eeqn
The minimum of the energy functional corresponds to the minimum of the new dimensionless parameter $\beta_{\rho}$. Contrary to the Abrikosov ratio~\eq{eq:beta}, the quantity~\eq{eq:beta:rho} depends on the strength of the magnetic field $B$. In the ``local'' (and, unphysical) limit $m_0 \to \infty$ the quantity~\eq{eq:beta:rho} is reduced to the Abrikosov ratio~\eq{eq:beta}.

We minimize numerically the mean energy density~\eq{eq:E:rho} as a function of (generally, complex) lattice parameters $C_n$, $n=1,\dots,N$ for a fixed value of $N=1,\dots,8$. We have found that the condensation energy,
\beqn
\delta {\cal E} = \langle {\cal E} \rangle - \frac{1}{2} B_\ext^2\,,
\label{eq:E:cond}
\eeqn
reaches its minimum at the equilateral triangular lattice with $N=2$ and $C_0 = i C_1$, similarly to the case of the Abrikosov lattice in the GL model. All lattices with odd values of $N$ possess higher energies while all even--$N$ lattices are reduced to the $N=2$ case. 

In Table~\ref{tbl:rho} we present the results of a numerical minimization of the energy functional~\eq{eq:E:rho} for fixed values of $N$ at $B=1.01 \,B_c$. As in the case of the GL model for a type-II superconductor, the difference in minimal condensation energy $\delta {\cal E}$ for the different values of $N$ is very small. The global minimum in $\delta {\cal E}$ and in the parameter $\beta_\rho$ is reached for even values of $N$ which corresponds to the same equilateral triangular lattice which is sometimes called ``hexagonal lattice''. Notice that the explicit parameterization of the parameters $C_n$ at different (even) values of $N$ may not correspond to 
multiple repeated copies of the $N=2$ solution (this fact is clearly seen for the case of $N=4$). The simplest square lattice is realized at $N=1$ while odd values of $N\geqslant 3$ correspond to a general (called ``oblique'') type of the 2 dimensional lattice.

Examples of the  minimal-energy lattices at $N=1,2,3,5$ formed in the magnetic field $B=1.01\, B_c$ are presented in Fig.~\ref{fig:rho:lattices}. The minimal-energy lattices with $N=4,6,\dots$ coincide with the $N=2$ solution.

\begin{figure}
\begin{center}
\includegraphics[width=80mm, angle=0]{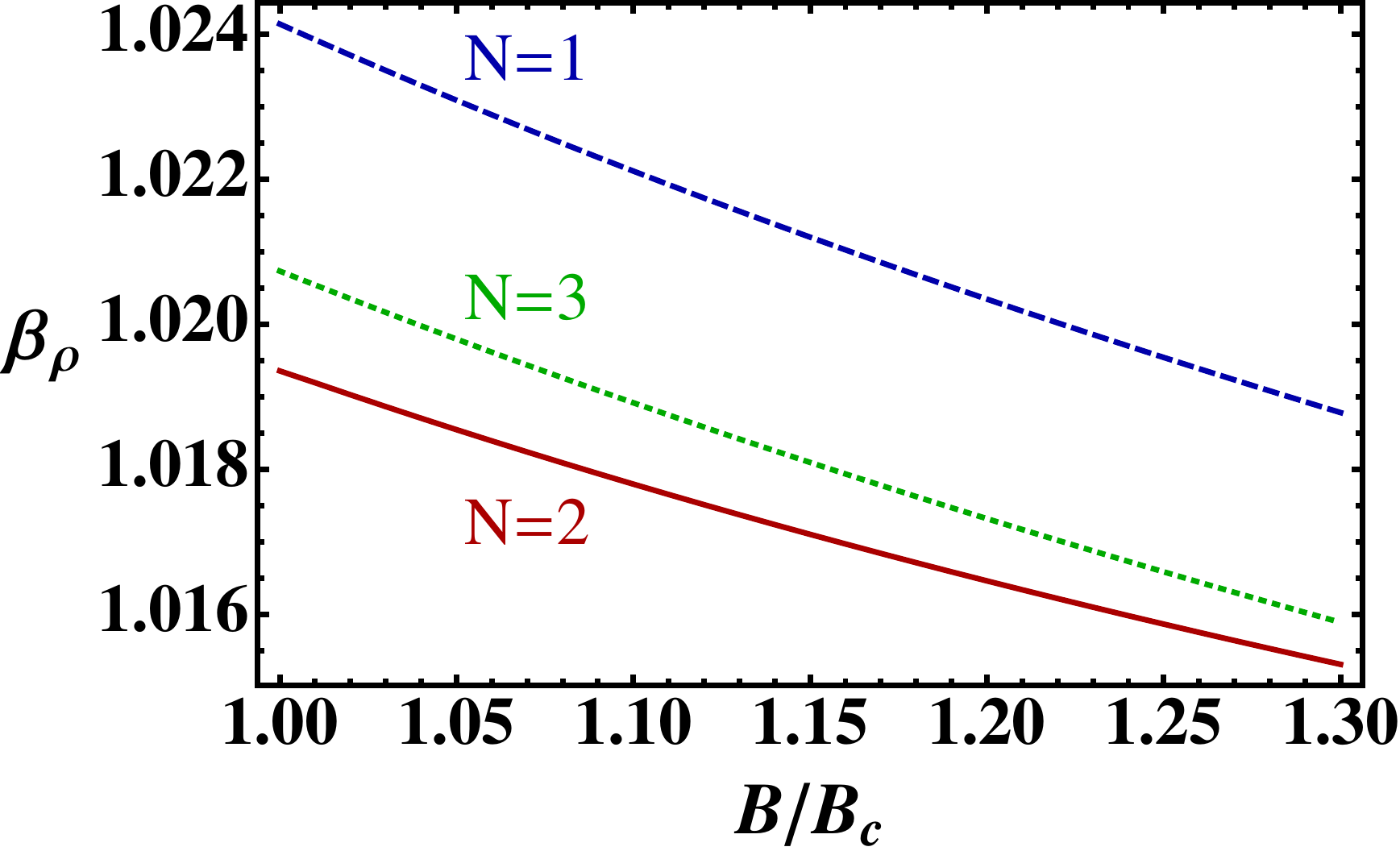}
\end{center}
  \caption{The parameter $\beta_\rho$, Eq.~\eq{eq:beta:rho}, for different types of lattices, $N=1,2,3$, as a function of the magnetic field~$B$.}
\label{fig:rho:beta}
\end{figure}
In Fig.~\ref{fig:rho:beta} we show the parameter $\beta_\rho$, Eq.~\eq{eq:beta:rho}, as a function of the magnetic field $B$ in the vicinity of the critical magnetic field $B_c$. The minimum of the quantity $\beta_\rho$ -- and, as a consequence, the minimum of the energy density~\eq{eq:E:rho} -- are reached at $N=2$ for all studied values of the magnetic field. Due to specific values of the phenomenological parameters of the DSGS model~\eq{eq:L:rho} -- which describes the electrodynamics of the $\rho$--meson excitations in the QCD vacuum -- the difference in energies between visually different lattices ({\it c.f.} Fig.~\ref{fig:rho:lattices}) is tiny. For example, at $B=1.01\, B_c$ the difference in the condensation energies between the square, $N=1$, lattice and the equilateral triangular, $N=2$, lattice is less than $0.5 \%$. The relative difference in the corresponding dimensionless $\beta$ parameters is of the same order.

In the first study of the vacuum superconductivity in Ref.~\cite{Chernodub:2010qx} the analysis of the ground state was done in assumption -- following the pioneering work of Abrikosov~\cite{Abrikosov:1956sx} -- that the vortex lattice has a square pattern with $N=1$. However, the more detailed analysis of this paper indicates that the real ground state of the superconducting vacuum is a triangular lattice\footnote{Soon after the Abrikosov's paper~\cite{Abrikosov:1956sx} the vortex pattern in a type-II superconductor was shown to be triangular~\cite{ref:Abrikosov}.} with $N=2$. Since the difference in most important bulk parameters (e.g., average energy, mean conductivity etc) between the square lattice and its possible conformations is very small, the square lattice is a very good approximation for calculation of the bulk properties of the real vacuum state. In this article we explore the correct (triangular vortex) state of the vacuum superconductivity which is important for local properties of this unusual phase.

\subsubsection{Superconducting condensate and energy density}

\begin{figure}
\begin{center}
\includegraphics[width=75mm, angle=0]{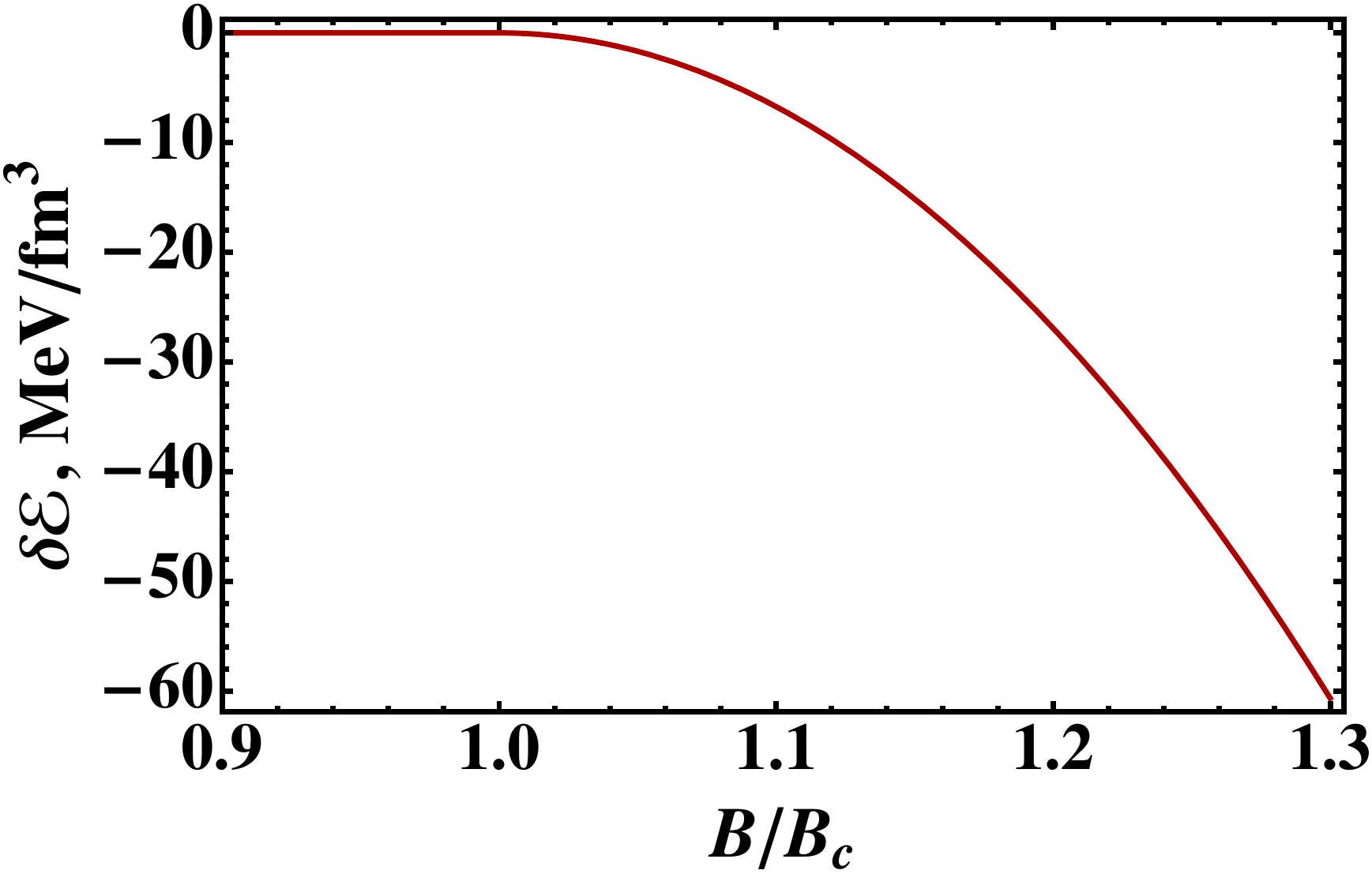}\\[5mm]
\hskip 4mm \includegraphics[width=71mm, angle=0]{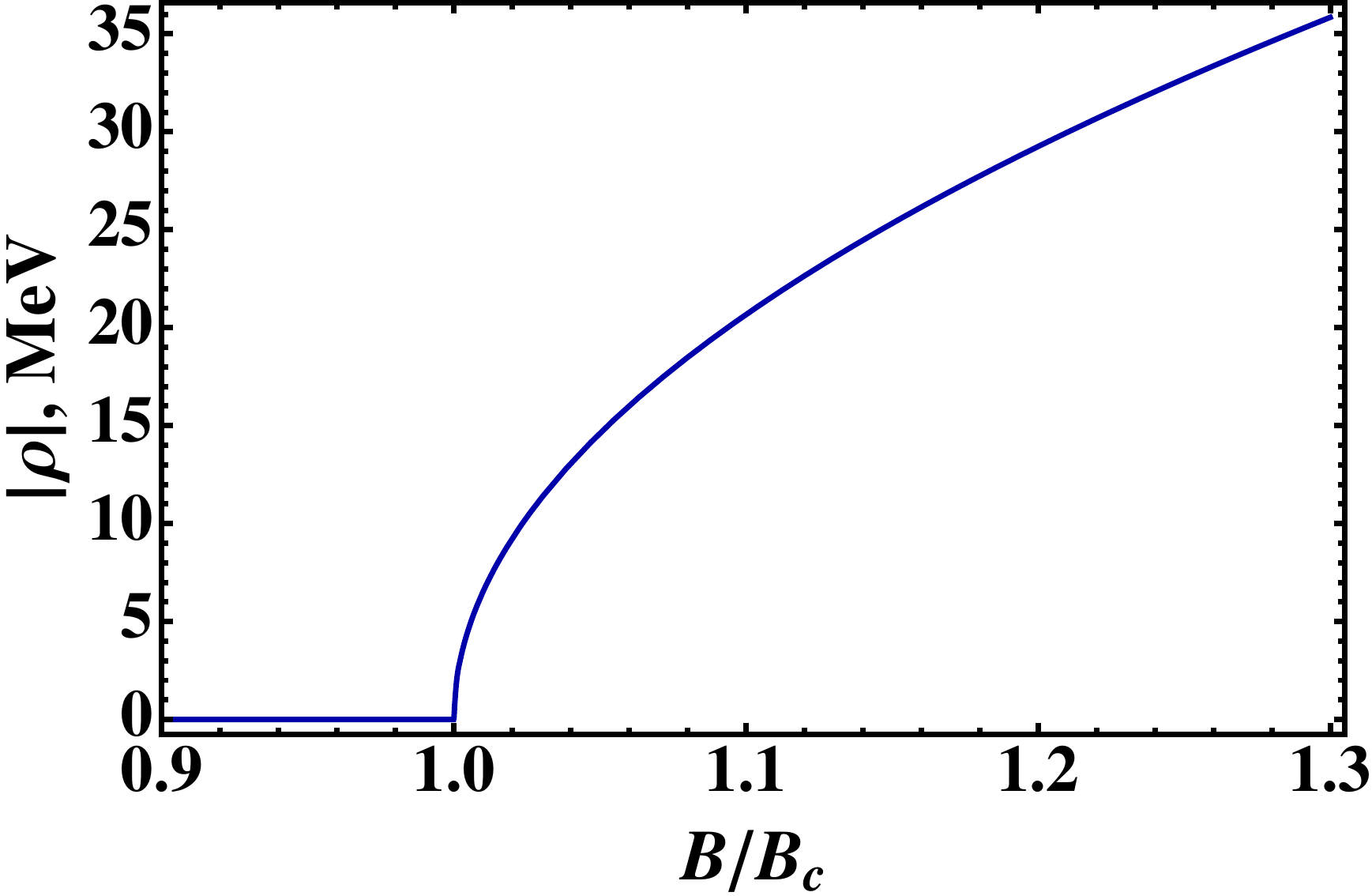}
\end{center}
\caption{At $B > B_c$ the superconducting state is more energetically favorable compared to the trivial vacuum state: (top) the condensation energy~\eq{eq:E:cond} becomes negative due to emergence of the superconducting condensate~\eq{eq:rho2}, $|\rho| \equiv \sqrt{\langle |\rho|^2 \rangle}$ (bottom) at the critical magnetic field $B=B_c$ with $B_c$ given in Eq.~\eq{eq:Bc}. The quantities are shown for the minimal-energy (equilateral triangular) lattice structure.}
\label{fig:rho:rho}
\end{figure}

From now on we concentrate on the triangular lattice which corresponds to two-fold symmetric ($N=2$) parameter space~\eq{eq:N:fold}, with the lowest-energy parameters given by Eq.~\eq{eq:N2:GL}.

In Fig.~\ref{fig:rho:rho} (top) we show the mean condensation energy density~\eq{eq:E:cond} as a function of the magnetic field. One can clearly see that at $B<B_c$ the condensation energy is zero while it becomes negative at $B>B_c$ due to condensation of the charged $\rho$ mesons. In order to characterize the later property, we notice that for the equilateral triangular vortex lattice the mean squared superconducting condensate is related to the coefficient $C_0$ of the solution~\eq{eq:rho:z} as follows:
\beqn
\langle |\rho|^2 \rangle \equiv \frac{1}{\mathrm{Area}_\perp} \int d^2 x \, |\rho(x)|^2= \frac{1}{2\sqrt[4]{3}} |C_0|^2\,.
\label{eq:rho2}
\eeqn

In Fig.~\ref{fig:rho:rho} (bottom) we plot the square root of the mean of the squared condensate~\eq{eq:rho2} as a function of the magnetic field. It is clear that at $B > B_c$ the superconducting state with a nontrivial condensate $\rho \neq 0$ is energetically more favorable compared to the trivial vacuum state with $\rho = 0$. The stronger magnetic field the larger the gain in energy due to the condensation of the $\rho$ mesons.
\begin{figure}[htb]
\begin{center}
\includegraphics[width=80mm, angle=0]{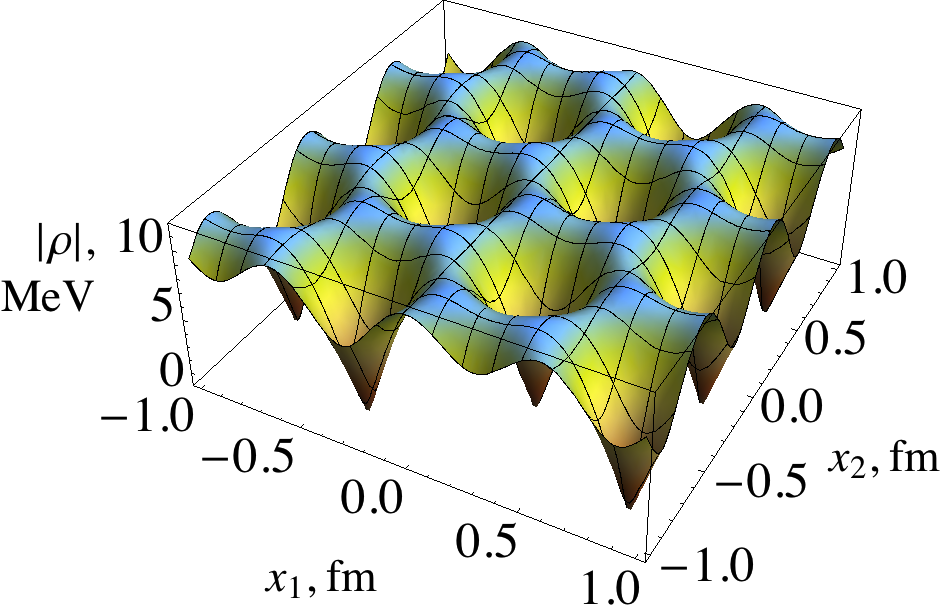}
\end{center}
  \caption{Absolute value of the superconducting condensate $\rho$, Eq.~\eq{eq:rho:z} at $B = 1.01 \, B_c$ in the transversal $(x_1,x_2)$ plane.}
\label{fig:rho}
\end{figure}
\begin{figure}[htb]
\begin{center}
\includegraphics[width=70mm, angle=0]{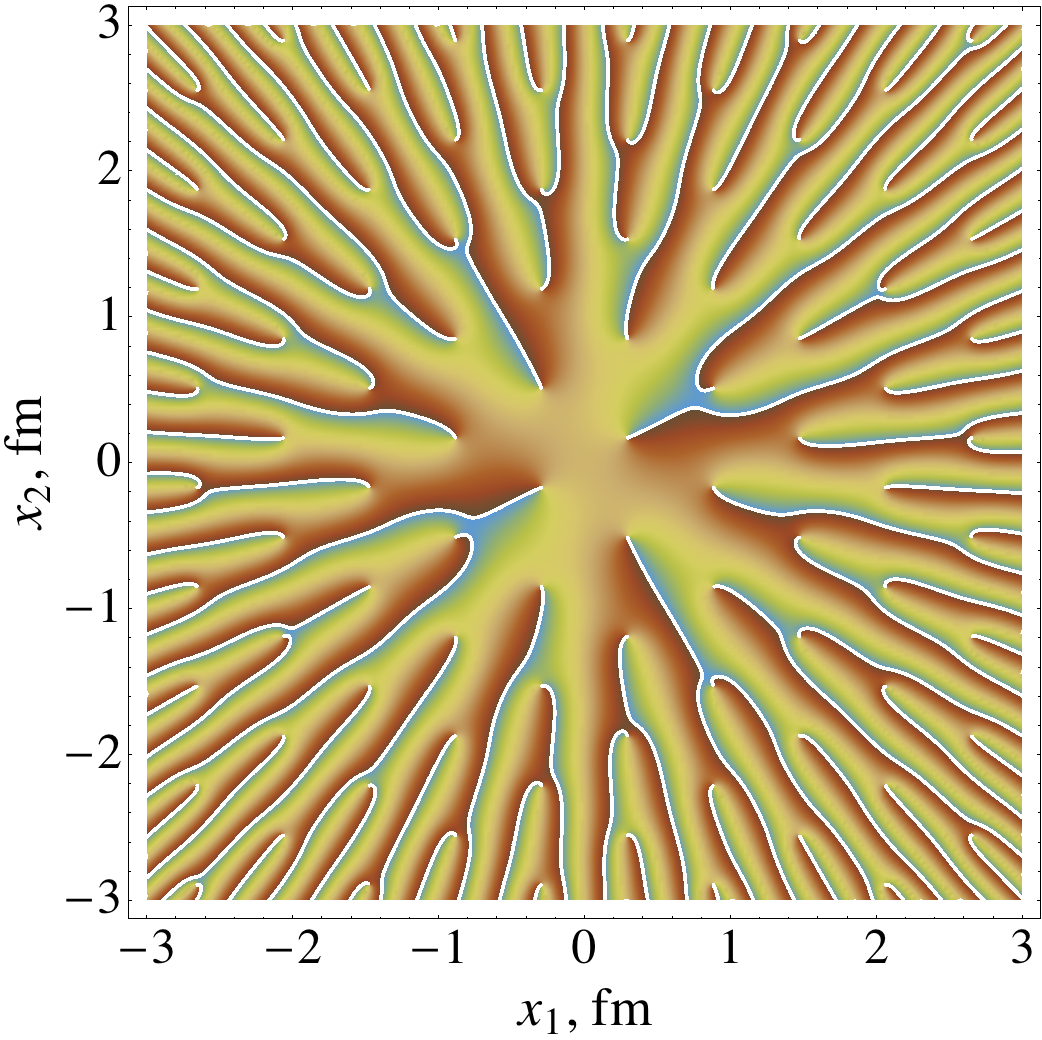}
\end{center}
  \caption{The density plot of the phase $\varphi_\rho = \arg \rho$ of the charged $\rho$--meson condensate~\eq{eq:rho:z} as a function of transversal coordinates $x_1$ and $x_2$. The white lines corresponds to the cuts in the phase, and the variations in color illustrate the behavior of the magnitude of the phases. The endpoints of the cuts mark the positions of the superconductor ($\rho$) vortices.  A three-fold difference in coordinate scales of this figure and Fig.~\ref{fig:rho} should be stressed.}
\label{fig:rho:cuts}
\end{figure}

The vortex structure of the superconducting ground state can easily be seen given the behavior of the superconducting order parameter $\rho$. The $\rho$ condensate has a characteristic form $\rho(z,\bar z) \propto z - z_0$ near the position $z_0 = x_{1,0} + i x_{2,0}$ of an elementary vortex. Thus, at the vortex core $z = z_0$ the condensate should vanish as a linear function. Moreover, in the local vicinity of the point $z_0$ the phase $\varphi_\rho = \arg \rho$ of the condensate should behave as a polar angle. Thus, the phase $\varphi_\rho$ winds around the position of the vortex and the winding corresponds to a topological stability of the vortex. Since the phase is defined modulo $2 \pi$, the phase $\varphi_\rho$ should experience cuts in the $(x_1,x_2)$ plane. At branches of these cuts the phase $\varphi_\rho$ experiences the quantized jumps $\varphi_\rho \to \varphi_\rho \pm 2 \pi$.

The absolute value and the phase of the superconducting condensate are shown in Fig.~\ref{fig:rho} and Fig.~\ref{fig:rho:cuts}, respectively. From Fig.~\ref{fig:rho} one can see that the condensate vanishes linearly ($\rho \propto |z - z_0|$) around a set of isolated points which form the equilateral triangular pattern, Fig.~\ref{fig:rho:lattices}(b). The phase of the condensate, Fig.~\ref{fig:rho:cuts}, experiences jumps at one-dimensional manifolds (semi-infinite lines) which start at the points where the condensate $\rho$ is vanishing, and end at spatial infinity. The position and the shape of the cuts can be changed by the $U(1)$ gauge transformations~\eq{eq:gauge:invariance}, while the endpoints of the cuts are gauge invariant quantities.

Thus, the $\rho=0$ points of the charged condensate do indeed mark the positions of the superconductor vortices\footnote{We call the topological defects in the charged $\rho$ condensate as ``superconductor vortices'' in order to distinguish them from ``superfluid vortices'' which are similar vortexlike defects in the neutral $\rho^{(0)}$ condensate.} and organize themselves into the equilateral triangle lattice.

\subsubsection{Electric currents and superconductivity}

The electric current density of the $\rho$ meson degrees of freedom can be derived from Eq.~\eq{eq:L:rho}:
\beqn
J_\mu & = & i e \bigl[\rho^{\nu\dagger} \rho_{\nu\mu} - \rho^\nu \rho^\dagger_{\nu\mu} + \partial^\nu (\rho^\dagger_\nu \rho_\mu - \rho^\dagger_\mu \rho_\nu)\bigr] \nonumber \\
& & - \frac{e}{g_s} \partial^\nu f^{(0)}_{\nu\mu}\,.
\label{eq:Jmu}
\eeqn
In the ground state the longitudinal components of the electric current are vanishing, $J_0 = J_3 = 0$, while the transversal current,
\beqn
J_\perp(x_\perp) \equiv J_1(x_\perp) + i J_2(x_\perp)\,,
\label{eq:J:complex}
\eeqn
is a nonlocal function of the superconducting condensate~\cite{Chernodub:2010qx}:
\beqn
J_\perp(x_\perp) = 2 i e m_0^2 \cdot \Bigl(\frac{\partial}{- \partial_\perp^2 + m_0^2} |\rho|^2\Bigr)(x_\perp) \,.
\label{eq:j:explicit}
\eeqn
The {\it nonlocal} nature of the relation between the transversal electric current~\eq{eq:j:explicit} and the charged condensate $\rho$ distinguishes the QCD vacuum from the GL superconductor~\eq{eq:L:GL}.

The electric current~\eq{eq:j:explicit} is a persistent current of the charged $\rho$-meson degrees of freedom. The current originates from the quarks and antiquarks which popup from the virtual state and form a condensate with the quantum numbers of the charged $\rho$-meson. This current is always present in the superconducting phase at $B > B_c$ and it vanishes in the normal phase of the vacuum. 

In order to simply further numerical calculations it is useful to represent the basic nonlocal part of Eq.~\eq{eq:j:explicit} 
\beqn
R(x_\perp;\rho) = \Bigl(\frac{m_0^2}{- \partial_\perp^2 + m_0^2} |\rho|^2\Bigr)(x_\perp)\,,
\label{eq:R}
\eeqn
as follows:
\beqn
R(x_\perp;\rho) = \int \frac{d^2 k}{(2 \pi)^2}  e^{i k_1 x_1 + i k_2 x_2} \frac{m_0^2}{k^2 + m_0^2} q(-k;\rho)\,, \quad
\eeqn
where the quantity $q(k;\rho)$ is defined in Eq.~\eq{eq:q:fun} and $\rho$ is given in Eq.~\eq{eq:rho:z} with $N=2$ symmetry. For the triangular lattice with $C_1 = i C_0$ one can explicitly show that 
\begin{widetext}
\beqn
& & q(k;\rho) = 2 \pi^2 |C_0|^2 \sum_{n_1\in \Z} \sum_{n_2\in \Z} \Biggl(
\Bigl[1 + (-1)^{n_2}\Bigr] \delta(\nu k_1 - 2 \pi n_1) \delta(k_2 - 2 \pi \nu n_2)
\exp\biggl[ - \frac{\pi}{2} \bigl(\nu^{-2} n_1^2 +\nu^2 n_2^2\bigr) \biggr]  \\
& &  \qquad +
\Bigl[1 - (-1)^{n_2}\Bigr] (-1)^{n_1 n_2} (-1)^{(n_2 + 1)/2} \delta\biggl[\nu k_1 - 2 \pi \Bigl(n_1 + \frac{1}{2}\Bigr) \biggr] \delta(k_2 - 2 \pi \nu n_2)
\exp\biggl\{ - \frac{\pi}{2} \Bigl[\nu^{-2} \Bigl(n_1 + \frac{1}{2}\Bigr)^2 +\nu^2 n_2^2\Bigr] \biggr\} 
\Biggr) \,, \nonumber
\eeqn
so that the nonlocal quantity~\eq{eq:R} is
\beqn
R(x_\perp;\rho) & = & \frac{|C_0|^2}{2 |\nu|} \sum_{n_1\in \Z} \sum_{n_2\in \Z} e^{-2 \pi i (n_1 x_1/\nu + n_2 x_2 \nu)}
\Biggl(\exp\biggl\{ - \frac{\pi}{2} \Bigl[\nu^{-2} n_1^2 +\nu^2 n_2^2\Bigr] \biggr\} 
\frac{\bigl[1 + (-1)^{n_2}\bigr]  \, \mu^2}{4 \pi^2 (\nu^{-2} n_1^2 + \nu^2 n_2^2) + \mu^2} \label{eq:R1} \\
& &  \qquad +
\exp\biggl\{ - \frac{\pi}{2} \Bigl[\nu^{-2} \Bigl(n_1 + \frac{1}{2}\Bigr)^2 +\nu^2 n_2^2\Bigr] \biggr\} 
\frac{\bigl[1 - (-1)^{n_2}\bigr] (-1)^{n_1 n_2} (-1)^{(1 - n_2)/2} \mu^2}{4 \pi^2 [\nu^{-2} (n_1+1/2)^2 + \nu^2 n_2^2] + \mu^2} e^{- \pi i x_1 / \nu}
\Biggr) \,, \nonumber
\eeqn
\end{widetext}
with $\mu = m_0 L_B$, where the magnetic length $L_B$ is given in Eq.~\eq{eq:LB}.  This expression is very convenient for numerical calculations because the sums in Eq.~\eq{eq:R1} converge very quickly. The derivatives of Eq.~\eq{eq:R1} with respect to the transverse coordinates $x_1$ and $x_2$ can be taken quite straightforwardly.

\begin{figure}
\begin{center}
\includegraphics[width=75mm, angle=0]{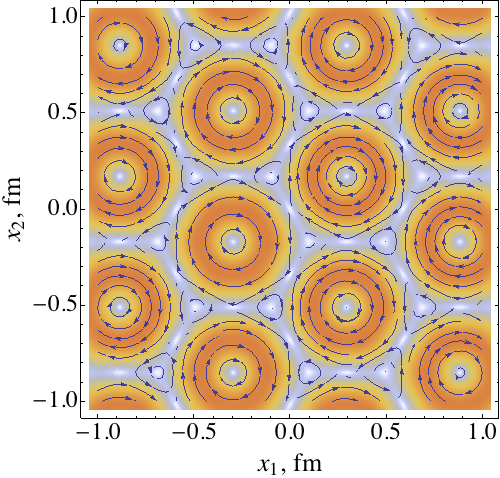}
\end{center}
\caption{The electric currents of the charged quark condensates in the ground state lattice at $B=1.01\, B_c$ in the transversal $(x_1,x_2)$ plane. The reddish (grayish) areas corresponds to the stronger (weaker) current.}
\label{fig:currents}
\end{figure}

In Fig.~\ref{fig:currents} we show the transverse electric current~\eq{eq:j:explicit} in the ground state of the superconducting vacuum at the magnetic field $B=1.01\, B_c$. All current lines are organized in the hexagonal pattern. The center of each hexagon corresponds to the position of a superconductor vortex.

\begin{figure}
\begin{center}
\includegraphics[width=80mm, angle=0]{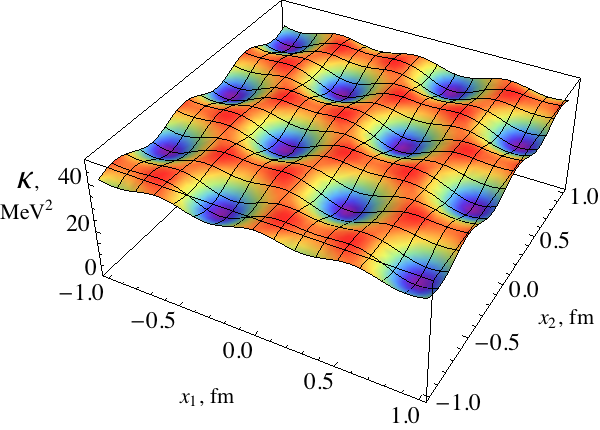}
\end{center}
  \caption{The superconductivity coefficient $\kappa$, Eq.~\eq{eq:kappa}, in the modified London law~\eq{eq:London}, is shown as a function of $x_1$ and $x_2$ transverse coordinates at the magnetic field $B=1.01\,B_c$.}
\label{fig:kappa}
\end{figure}

In transversal directions the strong magnetic field confines the local electric currents of charged condensates to the hexagonal cells. The size of cells are of the order of the lowest Landau level of the $\rho$-meson condensate, that is of the order of the magnetic length~\eq{eq:LB} ($L_B \sim 0.6\,{\mathrm{fm}}$ at $B \approx B_c$). A relatively week ($|\vec E| \ll |\vec B|$) electric field $\vec E \perp \vec B$ cannot create a transversal (``intra-cell'') electric current because such a current would involve an excitation at a first Landau level which is separated from a lowest Landau level by a large energy gap of the order of $E_1 - E_0 \sim \sqrt{|e B|}$. Thus, the global electric currents are drastically suppressed in the transverse directions. Therefore, in the transverse direction the vacuum state behaves as an insulator. This is the very reason why the Meissner effect is absent in the superconducting ground state~\cite{Chernodub:2010qx} so that the emerging superconductivity does not screen the external magnetic field.

However, the longitudinal electric currents can exist as the motion of the quarks along the axis of the magnetic field is not restricted. If one applies a weak electric field $\vec E = (0,0,E_3)$ parallel to the strong magnetic field $\vec B \equiv (0,0,B)$ then the electric currents -- induced by the external field -- satisfy a modified London equation~\cite{Chernodub:2010qx}:
\beqn
\frac{\partial J_3(x)}{\partial x_0}- \frac{\partial J_0(x)}{\partial x_3}  = - \kappa(x_\perp) E_3\,,
\label{eq:London}
\eeqn
while the {\it induced} transverse electric currents are obviously zero, so that the transverse current $J_\perp$ is unmodified by the external electric field directed along the magnetic field axis, Eqs.~\eq{eq:J:complex} and \eq{eq:j:explicit}, and
\beqn
\frac{\partial J_k(x)}{\partial x_\mu}- \frac{\partial J_\mu(x)}{\partial x_k}  \equiv 0 \,,
\label{eq:London:2}
\eeqn
with $\mu = 0,\dots,3$ and $k = 1,2$.

The London--type equation is typical for superconducting systems as it characterizes a conducting state without resistance. The superconductivity is characterized by the quantity $\kappa = \kappa(x_\perp)$, which is a nonlocal function of the superconducting condensate:
\beqn
\kappa(x_\perp) = 4 e^2 m_0^2 \cdot \Bigl( \frac{1}{- \partial_\perp^2 + m_0^2}  |\rho|^2\Bigr)(x_\perp)\,.
\label{eq:kappa}
\eeqn

In the superconducting ground state the ``superconducting transport coefficient'' $\kappa$ has a hexagonal lattice structure, Fig.~\ref{fig:kappa}. Due to nonlocal nature of the relation between the transport coefficient $\kappa$ and the superconducting condensate~\eq{eq:kappa}, the superconductivity is not completely suppressed inside the vortices. This property distinguishes the vacuum superconductivity from an ordinary one. In the GL model of superconductivity, the mentioned relation is local and the suppression of the condensation in the center of the Abrikosov vortex implies strong suppression of the superconductivity by the vortices.

Thus, at high magnetic field the vacuum becomes an anisotropic inhomogeneous superconductor. The superconductivity is anisotropic since the vacuum can superconduct along the direction of the magnetic field only and in the transversal directions the vacuum behaves as an insulator. The superconductivity is inhomogeneous due to the coordinate dependence of the superconducting transport coefficient $\kappa$, Eq.~\eq{eq:kappa}, Fig.~\ref{fig:kappa}.

\subsubsection{Neutral condensate and superfluid vortices}

It is known that the condensation of the charged (superconducting) field $\rho_\mu$ leads to the induced condensation of a neutral, superfluid-like field $\rho^{(0)}_\mu$ in the ground state of the vacuum~\cite{Chernodub:2010qx}. The longitudinal components of the neutral condensate are zero, $\rho^{(0)}_0 = \rho^{(0)}_3 = 0$, while the transverse components of the neutral condensate, are, in general, nonvanishing:
\beqn
\rho^{(0)}(x_\perp) = 2 i g_s \cdot \Bigl(\frac{\partial}{- \partial_\perp^2 + m_0^2} |\rho|^2\Bigr)(x_\perp) \,,
\label{eq:rho0:explicit}
\eeqn
where the quantity $\rho^{(0)} = \rho^{(0)}_1 + i \rho^{(0)}_2$ is a complex field.

The superfluid current $K_\mu$ -- which can be interpreted as a flow of the neutral condensate -- can be derived in a similar manner to the electromagnetic current~\eq{eq:Jmu} via variation of the action of the $\rho$--meson electrodynamics~\eq{eq:L:rho} with respect to the neutral field $\rho^{(0)}_\mu$. In the ground state of the vacuum the longitudinal superfluid current vanishes, $K_0=K_3 = 0$, while the transversal current,  
\beqn
K_\perp(x_\perp) \equiv K_1(x_\perp) + i K_2(x_\perp)\,,
\label{eq:K:complex}
\eeqn
is a nonlocal function of the superconducting condensate:
\beqn
K_\perp(x_\perp) = i \partial f_{12}^{(0)} \equiv 2 i g_s \cdot \Bigl(\frac{\partial_\perp^2 }{- \partial_\perp^2 + m_0^2} \partial |\rho|^2\Bigr)(x_\perp) \,. \quad
\label{eq:K:explicit}
\eeqn

\begin{figure}
\begin{center}
\includegraphics[width=80mm, angle=0]{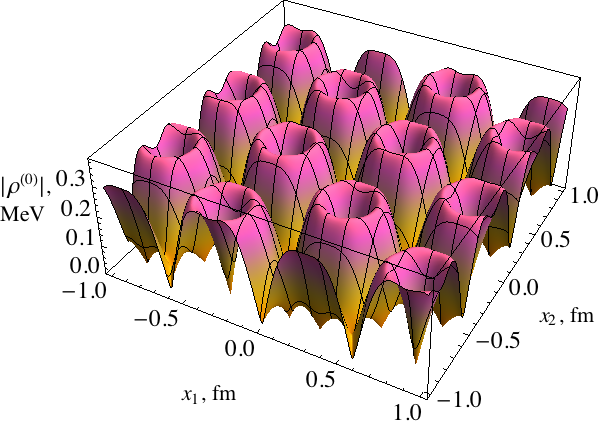}
\end{center}
  \caption{Absolute value of the superfluid condensate $\rho^{(0)}$, Eq.~\eq{eq:rho0:explicit}, at $B = 1.01 \, B_c$ in the transversal $(x_1,x_2)$ plane.}
\label{fig:rho0}
\end{figure}

\begin{figure}
\begin{center}
\includegraphics[width=80mm, angle=0]{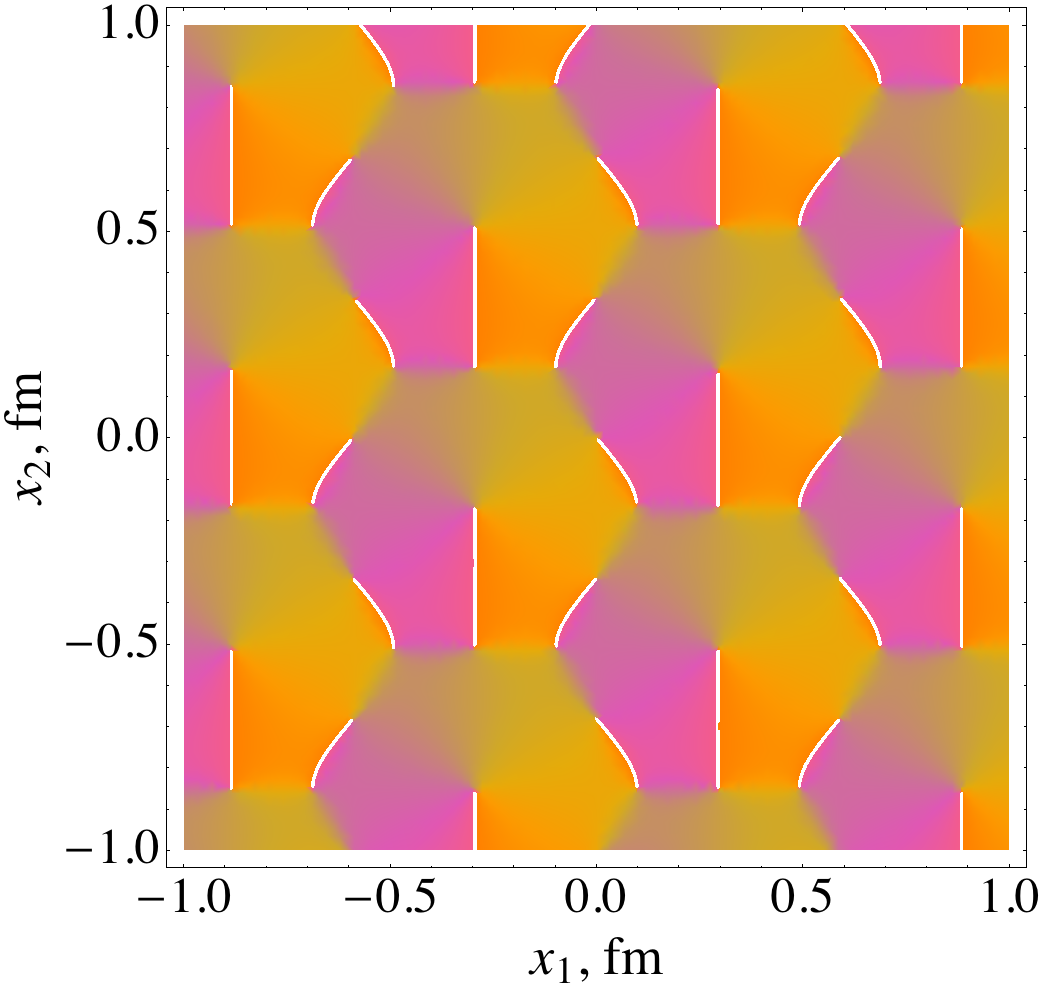}
\end{center}
  \caption{The density plot of the phase of the neutral $\rho$--meson field~\eq{eq:rho0:explicit} as a function of transverse coordinates $x_1$ and $x_2$. The white lines corresponds to the cuts in the phases, and the variations in color illustrate the behavior of the magnitude of the phase.}
  \label{fig:rho0:cuts}
\end{figure}

In Fig.~\ref{fig:rho0} we show the absolute value of the neutral condensates $\rho^{(0)}$ as a function of the transverse coordinates $x_1$ and $x_2$ for a slightly overcritical background magnetic field, $B = 1.01 \, B_c$.  One can compare this figure with Fig.~\ref{fig:rho} where the charged condensate is plotted for the same set of parameters. Firstly, we notice that the neutral condensate is much weaker compared to its charged counterpart, $|\rho^{(0)}| \ll |\rho|$. Secondly, the geometrical pattern of the neutral condensate is much more involved compared to the charged one. In particular one observes that the neutral $\rho$--meson condensate vanishes in a denser set of points compared to its superconducting counterpart. The latter set of points is organized in a triangular pattern, Fig.~\ref{fig:rho:lattices}. As the phases of the charged $\rho$ field are winding around these points (Fig.~\ref{fig:rho:cuts}), we concluded that these points correspond to the superconductor vortices (``$\rho$ vortices''). 
Following this analogy, one can suspect that the points of vanishing neutral condensate may correspond to ``superfluid'' vortices and antivortices. In order to figure out the topological structure of the neutral condensate we should check whether the phase of the neutral $\rho^{(0)}$ field winds around these points or not. 

In Fig.~\ref{fig:rho0:cuts} we show the density plot of the phase of the neutral $\rho$--meson field~\eq{eq:rho0:explicit} in the transverse plane $(x_1,x_2)$. The white lines corresponds to the cuts in the phase, so that the phase of the neutral condensate winds around the endpoints of these lines. At these endpoints the absolute value of the neutral field vanishes and the phase becomes undefined. Thus, the endpoints correspond to the (superfluid) vortices in the neutral $\rho$-meson field. 
Surprisingly, the superfluid vortices and antivortices comes always in pairs so that the net vorticity of the neutral field is zero.

\begin{figure}
\begin{center}
\includegraphics[width=80mm, angle=0]{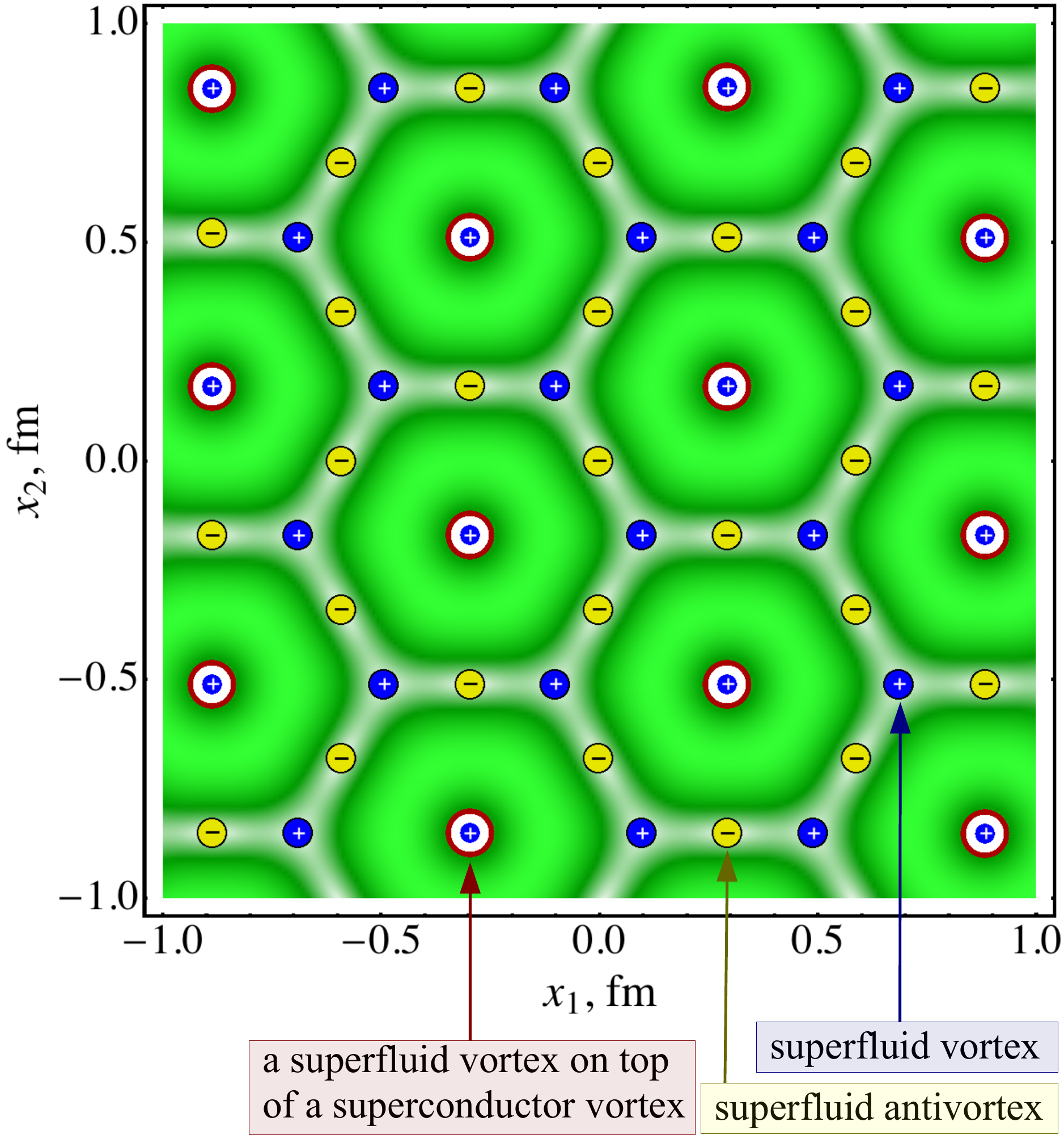}
\end{center}
  \caption{The periodic vortex structure of the vacuum ground state is superimposed on the density plot (shown in shades of the green color) of the absolute value of the neutral $\rho$--meson condensate~\eq{eq:rho0:explicit} at the magnetic field $B=1.01\,B_c$. Each superconductor vortex [the large red circles] is always superimposed on a superfluid vortex (the small blue disks marked by the plus signs) forming an equilateral triangular lattice. The isolated superfluid vortices and antivortices (the small yellow disks with the minus sign) arrange themselves in the hexagonal lattice pattern.}
\label{fig:rho:rho0}
\end{figure}

The superfluid current~\eq{eq:K:explicit} circulates around the superfluid vortices and antivortices in clockwise and counterclockwise in a manner which is visually similar to Fig.~\ref{fig:currents}. 

Finally, we would like to mention that the transversal electric current $J_\perp \equiv J_1 + i J_2$, Eq.~\eq{eq:j:explicit}, the superconducting transport coefficient $\kappa$, Eq.~\eq{eq:kappa}, the neutral condensate $\rho^{(0)} \equiv \rho^{(0)}_1 + i \rho^{(0)}_2$, Eq.~\eq{eq:rho0:explicit}, and the superfluid current $K_\perp$, Eq.~\eq{eq:K:explicit}, can be mutually related to each other:
\beqn
\frac{g_s}{e m_0^2} J_\perp (x_\perp) = \frac{i g_s}{4 e^2 m_0^2} \partial \kappa (x_\perp) & = & \rho^{(0)}(x_\perp) \,, \\
K_\perp (x_\perp) &  = & \partial_\perp^2  \rho^{(0)}(x_\perp)\,,
\eeqn
where $\partial \equiv \partial_1 + i \partial_2$ and $\partial_\perp^2 \equiv \partial_1^2 + \partial_2^2$ . This relation is valid in the leading order in the vicinity of the phase transition, $B \gtrsim B_c$.

\subsubsection{Superconductor and superfluid vortex lattices}

We show both the superconducting and superfluid vortices in Fig.~\ref{fig:rho:rho0}. The vortex locations are superimposed on the density plot of the absolute value of the neutral meson field~\eq{eq:rho0:explicit}. The vortex pattern is quite remarkable:
\begin{itemize}

\item The superconductor vortices organize themselves in an equilateral triangular lattice;

\item The superfluid vortices and antivortices are organized in a honeycomb (hexagonal tiling) pattern;

\item A center of each hexagon is  occupied by a superconductor vortex;

\item Each superconductor vortex is always superimposed on a superfluid vortex;

\item One superconductor vortex is accompanied by three superfluid vortices and three superfluid antivortices so that the net superfluid vorticity is zero.

\end{itemize}

\section{Conclusion}

In this paper we determine the structure of inhomogeneities of the quantum vacuum in a superconducting state. The superconducting state of the vacuum is realized at a strong magnetic field background if the strength of the field exceeds the critical value~\eq{eq:Bc}, Ref.~\cite{Chernodub:2010qx}. The superconductivity is associated with a spontaneous emergence of quark-antiquark condensates which carry quantum numbers of charged $\rho$ mesons. The $\rho$--meson condensate in the ground state turns out to be an inhomogeneous structure made of the vortexlike defects in the $\rho$-meson field. These superconductor vortices are parallel to the magnetic field axis, and in this paper we show that in the transversal plane they organize themselves in a equilateral triangular lattice. 

The condensation of the charged $\rho$ mesons induces a (much weaker) superfluid-like condensate with quantum numbers of the neutral $\rho^{(0)}$ mesons~\cite{Chernodub:2010qx}.  In this paper we have shown that the neutral condensate contains vortexlike defects as well (by analogy we call them as the ``superfluid vortices''). We show that each of these superconductor vortices is accompanied by three superfluid vortices and three superfluid antivortices made of the neutral $\rho$ meson condensate. Moreover, one of these superfluid vortices overlaps with a superconductor vortex, so that each superconductor vortex is always accompanied by the superfluid vortex. The superfluid vortices organize themselves in a honeycomb lattice.

\acknowledgments

The work of MNC was partially supported by Grant No. ANR-10-JCJC-0408 HYPERMAG.

\end{document}